\documentclass[pra,twocolumn,a4paper,superscriptaddress,nofootinbib,showpacs]{revtex4}
\usepackage{amsmath}
\usepackage{amsfonts}
\usepackage{color}
\usepackage{graphicx}

\begin{document}

\title{Quantum communication using a bounded-size quantum reference frame}
\author{Stephen D. Bartlett}
\affiliation{School of Physics, The University of Sydney, Sydney, New South Wales 2006, Australia}%
\author{Terry Rudolph}%
\affiliation{Optics Section, Blackett Laboratory, Imperial College London, London SW7 2BZ, United Kingdom}%
\affiliation{Institute for Mathematical Sciences, Imperial College London, London SW7 2BW, United Kingdom}%
\author{Robert W. Spekkens}%
\affiliation{Perimeter Institute for Theoretical Physics, 31 Caroline
St. North, Waterloo,
  Ontario N2L 2Y5, Canada}
\author{Peter S. Turner}%
\affiliation{Department of Physics, Graduate School of Science, University of Tokyo, Tokyo 113-0033, Japan}%
\date{3 April 2009}

\begin{abstract}
Typical quantum communication schemes are such that to achieve
perfect decoding the receiver must share a reference frame with the
sender.  Indeed, if the receiver only possesses a bounded-size
quantum token of the sender's reference frame, then the decoding is
imperfect, and we can describe this effect as a noisy quantum
channel. We seek here to characterize the performance of such
schemes, or equivalently, to determine the effective decoherence
induced by having a bounded-size reference frame. We assume that the
token is prepared in a special state that has particularly nice
group-theoretic properties and that is near-optimal for transmitting
information about the sender's frame.  We present a decoding
operation, which can be proven to be near-optimal in this case, and
we demonstrate that there are two distinct ways of implementing it
(corresponding to two distinct Kraus decompositions).  In one, the
receiver measures the orientation of the reference frame token and
reorients the system appropriately. In the other, the receiver
extracts the encoded information from the virtual subsystems that
describe the relational degrees of freedom of the system and token.
Finally, we provide explicit characterizations of these decoding
schemes when the system is a single qubit and for three standard
kinds of reference frame: a phase reference, a Cartesian frame
(representing an orthogonal triad of spatial directions), and a
reference direction (representing a single spatial direction).

\end{abstract}

 \maketitle

\section{Introduction}

Many communication protocols implicitly require that the communicating parties share a reference
frame~\cite{BRSreview}.  For instance, if one party, Alice, transmits qubits to another party, Bob, using
spin-1/2 particles, the quantum state of the collection can only be recovered by Bob if he and Alice share a
reference frame for orientation.  Lacking such a shared reference frame (for instance, by lacking knowledge of
the relation between their local reference frames) is equivalent to having a noisy channel; the density operator
relative to Bob's local frame is the average over rotations of the density operator relative to Alice's local
frame.  For a single qubit, such an average over rotations yields complete decoherence -- no information about
the quantum state survives. Nonetheless, Alice and Bob can still achieve perfect classical and quantum
communication by encoding the information into the rotationally-invariant degrees of freedom of many
qubits~\cite{BRS03}. Indeed, in the limit of large numbers, the cost of not sharing a reference frame is only
logarithmic in the number of systems. Similarly, if Alice and Bob lack a phase reference, they can still encode
classical and quantum information in phase-invariant states of composite systems~\cite{BRSreview}. However, such
communication schemes are technically challenging to implement because they make use of highly entangled states
of many qubits~\cite{Ban04,Bou04}.  Such schemes will be referred to here as ``calibration-free''.

\begin{figure}[!]
\begin{center}
\includegraphics[width=80mm]{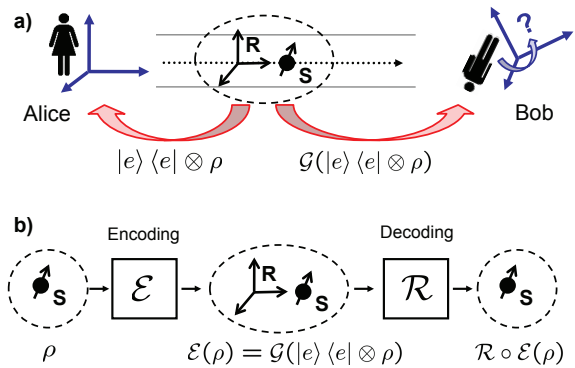}
   \caption{a) Alice describes the reference frame token $R$ by the state $|e\rangle\langle e|$ and the system $S$ by the state $\rho$.
   Bob, who is not correlated with Alice's reference frame, describes the joint state of $R$ and $S$ as the twirling of Alice's description,
   namely $\mathcal{G} (|e\rangle \langle e| \otimes \rho )$.  b) We can consider Alice's act of adjoining the reference frame token to the
   system together with the twirling as an encoding operation $\mathcal{E}$. Composing this with Bob's decoding operation $R$, the resulting
   channel can be described as an effective $\mathcal{R} \circ \mathcal{E}$.}\label{fig:figure1}
\end{center}
\end{figure}

A more straightforward strategy for coping with the lack of a shared
reference frame is for Alice and Bob to begin their communication
protocol by setting up a shared reference frame, that is, they begin
by calibrating or aligning their local reference frames. Thereafter
Alice transmits her quantum systems normally.  This strategy is
illustrated in figure~\ref{fig:figure1}. The problem of aligning
reference frames using finite communication resources has been well
studied~\cite{Gis99,Per01a,Per01b,Bag04,Chi04,BRSreview}. If only
finite communication resources are devoted to the task, then the
token of Alice's reference frame that is transmitted to Bob will be
of bounded size. Assuming the reference frame in question is
associated with a continuous degree of freedom, this bound leads to
a nonzero probability of error in the decoding of messages.
Nonetheless, this scheme has two advantages over its
calibration-free counterpart: (i) Alice does not need to implement
any entangling operations to encode classical bits or logical qubits
into the physical qubits, nor does Bob require such operations to
decode (They may require entangling operations to prepare and
measure the reference frame token, but the preparation and
measurement they require is always the same and their effort does
not scale with the size of the message); (ii) If Alice wishes to
communicate a classical bit string or a string of logical qubit
states, she can encode one logical bit or qubit per physical qubit,
and Bob can decode one logical bit or qubit per physical qubit (in
other words, the blocks of physical qubits into which they encode
and decode their logical bits and qubits can be as small as one,
unlike the calibration-free scheme).  By virtue of (ii), Alice does
not need to know the entire message string at the outset to achieve
her optimal communication rate, nor does Bob need to store all of
the systems coherently until he has received the entire sequence of
physical qubits, whereas such capabilities are required to achieve
the optimal rate of communication in the calibration-free scheme. We
are therefore motivated to explore how well Alice can communicate
quantum information to Bob after supplying him with a bounded-size
token of her reference frame.

The general case we consider is that of a reference frame associated
with a compact Lie group $G$. If the system is prepared in the state
$\rho$ and the reference frame token in the state $|e\rangle$
relative to Alice's local reference frame, then relative to Bob's
local reference frame, the pair is in the state
\begin{equation} \label{eq:effectivedecoherence}
  \mathcal{E}[\rho] = \mathcal{G} (|e\rangle \langle e| \otimes \rho )\,,
\end{equation}
where $\mathcal{G}$ averages over the collective action of the group
and is termed the $G$-twirling operation (see
figure~\ref{fig:figure1}). This state encodes $\rho$. We consider
the case where the reference frame token is prepared in a particular
state $|e\rangle$, suggested by previous
investigations~\cite{KMP04,Chi04}, which is near-optimal for
transmitting information about the group element and which makes the
mathematics particularly simple.  We then determine how well Bob can
reconstruct the state $\rho$.  It turns out that the recovery (or
decoding) operation $\mathcal{R}$ that is equal to the
Hilbert-Schmidt adjoint of $\mathcal{E}$, normalized to be
trace-preserving, is provably near-optimal (in a sense we will
define later).  The composition of encoding and decoding yields an
effective decoherence of the form
\begin{equation} \label{eq:effectivedecoherence2}
  (\mathcal{R}\circ\mathcal{E})(\rho) = \int \text{d}g\, p(g) U_S(g)\rho U_S^{\dag }(g)\,,
\end{equation}
where
\begin{equation}
  p(g) \propto |\langle e|U_R(g)|e\rangle|^{2} \,,
\end{equation}
is a probability distribution over the group with $\text{d}g$ the
group-invariant measure, $U_S$ and $U_R$ are unitary representations
of $G$ on the system $S$ and the reference frame token $R$
respectively.  With our particular choice of reference state
$|e\rangle$, we find that $p(hgh^{-1}) =p(g)$ for all $h\in G$,
ensuring that $\mathcal{R}\circ\mathcal{E}$ is a $G$-invariant map
(it commutes with the action of every $g\in G$). The figure of merit
relative to which the decoding operation is judged to be
near-optimal is the entanglement fidelity.

We also demonstrate two distinct ways of implementing this decoding
operation.  The first is an obvious scheme: Bob estimates the
orientation of the token relative to his local frame and then
re-orients the system appropriately. We call this the ``measure and
re-orient'' scheme. The second is less intuitive, but reveals more
about the structure of the problem: Bob projects into the virtual
subsystems that support the representation of the group induced by
\emph{relative} transformations of the system and token and
implements an isometry that maps these onto a single Hilbert space.
We call this the ``extract from the relational subsystems'' scheme.
These relational subsystems are the places in the Hilbert space of
the combined token and system where the quantum information
associated with $\rho$ is to be found. Their characterization is the
key technical result of the article. The second scheme is also
particularly interesting because, with a slight modification, it can
yield a decoding that is probabilistically perfect, that is, one
which sometimes fails but which yields a perfect decoding when it
succeeds.

We work out the explicit form of the recovery operation
$\mathcal{R}$ and the effective decoherence
$\mathcal{R}\circ\mathcal{E}$ for several interesting examples: (i)
a phase reference; (ii) a Cartesian frame (representing an
orthogonal triad of spatial directions)\footnote{Note that although
the term ``Cartesian frame'' is commonly used to refer to a
reference frame for \emph{both} orientation of the axes of an object
as well as the object's position, it is here used only for
orientation.}; (iii) a reference direction (representing a single
spatial direction). In each case, we consider a system consisting of
a single qubit. The explicit form of the decoherence map is actually
quite simple in this case. Because $\mathcal{R}\circ\mathcal{E}$ is
a $G$-invariant map, it follows from the results of
Ref.~\cite{Boi08} that it is a sum of irreducible $G$-invariant maps
called moments.  But in the case of a qubit, there is only a single
nontrivial moment -- the $G$-twirling map $\mathcal{G}$. Thus, we
have
\begin{equation}
  \mathcal{R}\circ\mathcal{E} =(1-p)\mathcal{I}+p\mathcal{G}\,,
\end{equation}
where $\mathcal{I}$ is the identity map.
We show that for our examples, $p$ is inversely proportional to the size of the reference frame token,
\begin{equation}
  p \propto \frac{1}{\textrm{size of RF}}\,,
\end{equation}
where the size of the reference frame is given by the quantum number of the highest irreducible representation
appearing in the state of the token.  It is also shown that in probabilistically perfect decoding schemes, the
probability of failure is inversely proportional to the size of the reference frame token.  These results are
specific to the special form of reference frame state that we consider here.

The idea that bounded-size quantum reference frames induce an effective decoherence is not new. The effect of a
bounded-size clock, which is a kind of phase reference, has been considered previously by many authors
\cite{Gam04,Gam04b,Mil06,Pag83} and that of a bounded-size directional frame has been considered by
Poulin~\cite{Pou06}.  Our results go beyond this work in several ways.  For one, the case of a Cartesian frame,
which is particularly significant given that it is representative of the general non-Abelian case, has not been
examined before.
More significantly, we consider many different sorts of reference
frames within a unified framework and we provide insight into the
structure of the problem.  It should be noted that although our
method can be applied to a system of arbitrary dimension, we here
obtain explicit expressions for the effective decoherence only in
the case of the simplest possible system: a qubit. We hope to
provide a discussion of the foundational implications of modeling
bounded-size reference frames by effective decoherence in a
subsequent paper.

Finally, it is worth pointing out that the problem of communication
in the presence of a bounded-size reference frame has interesting
connections with a disparate set of topics in quantum information
theory and the theory of quantum reference frames:

\emph{Partially correlated reference frames.} When some information is known about the relative orientation of
Alice and Bob's local reference frames, they are said to share partially correlated reference frames. This is a
resource that interpolates between having and lacking a shared reference frame.  Its quality can be
characterized by the probability distribution over the relative orientation -- the more peaked the distribution,
the better the correlation. What can be achieved with this resource is an interesting question that has only
begun to be addressed.  We gain some insight into the question in this article because the ``measure and
re-orient'' implementation of the recovery operation begins with a reference frame alignment
protocol~\cite{Gis99,Per01a,Per01b,Bag04,Chi04,BRSreview} that leaves Alice and Bob holding partially correlated
reference frames.


\emph{Programmable operations.} There have been many investigations into the possibility of encoding an
operation into a quantum state such that the state can subsequently be used to implement the operation on
another system.  If the system into which the operation is encoded is bounded in size, then it is known that one
can only achieve an approximate version of the operation or a perfect version with non-unit
probability~\cite{VC00,Dus02,Fiu02}. The token of the quantum reference frame in our communication protocol is
an instance of a program that encodes the unitary that relates Alice's local reference frame to Bob's. Bob
subsequently uses it to implement the inverse of this unitary on the system.  Our results therefore provide
interesting examples of both approximate and unambiguous programmable operations.

\emph{Measures of the quality of a quantum reference frame.} Our results also contribute to the project of
quantifying the extent to which a quantum reference frame of bounded size can stand in for one of unbounded
size~\cite{BRST06,BRST07,Gou08}, an important element of a resource theory of quantum reference frames. For
instance, a measure of the strength of the effective decoherence associated with the bounded-size reference
frame may serve as an operational measure of its quality.

\emph{Private channels.} If no party besides Bob has a sample of Alice's RF, then Alice and Bob are said to
possess a \emph{private} shared reference frame. These have been shown to constitute a novel kind of key that is
useful for private communication schemes~\cite{BRS04b}. Our results establish a lower bound on the fidelity
between input and output in private communication schemes that rely on Bob possessing a bounded-size token of
Alice's reference frame. They also establish a bound on the probability of achieving perfect fidelity by
post-selection. This cryptographic application provides a particularly useful perspective on Bob's recovery
operation: the message is encoded in the relative orientation between the system and the token of Alice's
reference frame in the same way that the plain-text in classical cryptography is encoded in the bit-wise parity
of the cypher-text message and the key. Bob decodes by using the token of Alice's frame as a key.


\emph{Dense coding.} As noted above in point (ii), the
calibration-free communication scheme requires Alice to know the
entire message to be sent prior to transmitting any systems to Bob
if she is to achieve the maximum rate. In the communication scheme
that first sets up a bounded-size reference frame, Alice can
transmit to Bob the quantum token of her local reference frame prior
to knowing anything about the message. Consequently, Alice can use
the quantum channel at an early time, when it is perhaps cheaper, to
enhance the communication capacity at a future time, when it is
known what message is to be sent. Sending the frame token is
therefore akin to establishing entanglement in a dense coding
scheme~\cite{Ben92}. One difference, however, is that the fidelity
of communication is never perfect for a bounded-size token whereas a
single maximally-entangled state allows for perfect communication of
one qubit.

\subsection{Mathematical preliminaries}

In this section, we present some formal mathematical tools that are useful for describing classical and quantum
reference frames.  We follow the notation of~\cite{BRSreview}, to which we refer the reader for further details.
Suppose Alice and Bob are considering a single quantum system described by a Hilbert space $\mathcal{H}$.  Let
this system transform via a representation of a group $G$ relative to some reference frame. We will restrict our
attention to Lie groups that are compact, so that they possess a group-invariant (Haar) measure
$\text{d}g$, and act on $\mathcal{H}$ via a \emph{unitary} representation $U$, ensuring that they
are completely reducible~\cite{Sternberg}.

Let $g\in G$ be the group element that describes the passive
transformation from Alice's to Bob's reference frame. Furthermore,
suppose that $g$ is completely unknown, i.e., that Alice's reference
frame and Bob's are uncorrelated.  It follows that if Alice prepares
a system in the state $\rho$ on $\mathcal{H}$ relative to her frame,
then it is represented relative to Bob's frame by the state
\begin{equation}
   \mathcal{G}[\rho]  = \int_G \text{d}g\, U(g) \rho U^\dag(g)   \, .
  \label{eq:AveragedState}
\end{equation}
The action of the representation $U$ of the (compact Lie) group $G$ on $\mathcal{H}$ yields a very useful
structure. It allows for a decomposition of the Hilbert space into a direct sum of \emph{charge sectors},
labeled by an index $q$, where each charge sector carries an inequivalent representation $U^{(q)}$ of $G$.  Each
sector can be further decomposed into a tensor product of a subsystem $\mathcal{M}^{(q)}$ carrying an
irreducible representation (irrep) of $G$ and a subsystem $\mathcal{N}^{(q)}$ carrying a trivial
representation of $G$.  That is,
\begin{equation}
    \label{eq:HdecompFull}
  \mathcal{H} = \bigoplus_q \mathcal{M}^{(q)} \otimes \mathcal{N}^{(q)} \,.
\end{equation}
Note that this tensor product does not correspond to the standard tensor product obtained by combining multiple
qubits: it is \emph{virtual}~\cite{Zan04}.  The spaces $\mathcal{M}^{(q)}$ and $\mathcal{N}^{(q)}$ are
therefore \emph{virtual subsystems}.

Expressed in terms of this decomposition of the Hilbert space, the
map $\mathcal{G}$ takes a particularly simple form, given by
\begin{equation}
  \label{eq:GeneralDecoherenceDfullDfree}
  \mathcal{G}[\rho] = \sum_q (\mathcal{D}_{\mathcal{M}^{(q)}} \otimes
  \mathcal{I}_{\mathcal{N}^{(q)}})[\Pi^{(q)} \rho \Pi^{(q)}] \,,
\end{equation}
where $\Pi^{(q)}$ is the projection into the charge sector $q$, $\mathcal{D}_{\mathcal{M}}$ denotes the
trace-preserving operation that takes every operator on the Hilbert space $\mathcal{M}$ to a constant times the
identity operator on that space, and $\mathcal{I}_{\mathcal{N}}$ denotes the identity map over operators
in the space $\mathcal{N}$.  A proof of this result is provided in Ref.~\cite{BRSreview}.

Note that the operation $\mathcal{G}$ has the general form of
\emph{decoherence}.  Whereas decoherence typically describes
correlation with an environment to which one does not have access,
in this case the decoherence describes correlation with a reference
frame to which one does not have access~\cite{BRS04a}.

\section{Encoding}

Consider a communication scheme wherein Alice prepares a system $R$
in a pure quantum state $|e\rangle$ and sends it to Bob as a quantum
sample of her reference frame, together with a system $S$ (a
collection of qubits for example) that is described by a quantum
state $\rho$ relative to her reference frame.  Let $R$ transform via
the unitary representation $U_R$ of $G$, and $S$ via the unitary
representation $U_S$.  The lack of a shared reference frame between
Alice and Bob implies that the transmitted composite $RS$ is
described relative to Bob's reference frame by the $G$-invariant
state
\begin{equation}
  \mathcal{E}(\rho) = \mathcal{G}_{RS}[|e\rangle\langle e| \otimes \rho] \,,
  \label{eq:GInvState}
\end{equation}
where $\mathcal{G}_{RS}$ is the $G$-twirling operation
of~(\ref{eq:AveragedState}) for the representation $U_{RS} = U_R
\otimes U_S$ of $G$. This map $\mathcal{E}$ will be referred to as
the \emph{encoding map}. Note that its input space is
$\mathcal{B}(\mathcal{H}_S)$ (the bounded operators on
$\mathcal{H}_S$), while its output space is
$\mathcal{B}(\mathcal{H}_R \otimes \mathcal{H}_S)$.  It maps $\rho$
to a $G$-invariant state of the composite $RS$.

It is useful to define the set of states
\begin{equation}
  \label{eq:CovariantRFStates}
 \{  |g\rangle= U_R(g)|e\rangle | g \in G \}\, ,
\end{equation}
which form the orbit under the representation $U_R$ of $G$ of the fiducial state $|e\rangle$ (associated with
the identity element of the group).  By expressing the $G$-twirling operation explicitly and making use of
these, we can express the encoding operation as
\begin{equation}
  \label{eq:Encoding}
  \mathcal{E}(\rho) = \int \text{d}g\, |g\rangle \langle g| \otimes U_S(g)\rho U_S^{\dag }(g) \,.
\end{equation}
The encoding map clearly depends on the choice of the state
$|e\rangle$ for the reference frame. We turn to this choice now.


\subsection{Quantum reference frames}

We begin by considering what properties of a quantum state make it a good representative of a reference frame
for the group $G$ (the case wherein the reference frame is associated with a coset space will be considered in
Section~\ref{sec:Direction}); for a more complete discussion, see~\cite{BRSreview,KMP04}.

States $|g\rangle$ corresponding to different orientations of the reference frame must be distinct, so at the
very least one requires that the the fiducial state $|e\rangle$ is not invariant with respect to $G$ or any
subgroup thereof. To emulate a perfect reference frame for $G$, these states must in fact be perfectly
distinguishable,
\begin{equation}
  \label{eq:DeltaFunction}
  \langle g|g'\rangle = \delta(g^{-1}g')\,,
\end{equation}
where $\delta(g)$ is the delta-function on $G$ defined by $\int
\mathrm{d}g\,\delta(g)f(g)=f(e)$ for any continuous function $f$ of
$G$, where $e$ is the identity element in $G$.  If the states $\{
|g\rangle \}$ of Eq.~(\ref{eq:CovariantRFStates}) satisfy these
requirements, then $U_R$ is the \emph{left regular representation}
of $G$.  In the case of a Lie group, the dimensionality of any
system $\mathcal{H}_{R}$ that carries the regular representation
must necessarily be infinite. We refer to such an
infinite-dimensional quantum RF as \emph{unbounded}; such systems
and states were considered in Ref.~\cite{KMP04}.

If the Hilbert space dimensionality of the system $R$ serving as a quantum RF is finite, then we say that
the quantum RF is \emph{of bounded size}.   If the RF is associated with a Lie group, having a continuum of
elements, then a bound on the size of the RF implies that the condition (\ref{eq:DeltaFunction}) cannot be
satisfied precisely. In this case, a key question is: what state on $R$ is the best approximation to a perfect
reference frame? The answer will depend on the figure of merit for the task at hand, but we will make use of a
generic construction~\cite{KMP04,Chi04} that illustrates the key features.

Suppose the representation $U_R$ reduces to a set of irreps $\{ U_R^{(q)} \}$,
\begin{equation}
  U_R(g) = \bigoplus_{q} U^{(q)}_R(g) \otimes I\,,
\end{equation}
where the tensor product is the one appearing in the decomposition
(\ref{eq:HdecompFull}) of $\mathcal{H}_R$ and where $I$ is the
identity on $\mathcal{N}^{(q)}_R$. We are interested in a special
subset of these irreps, namely, the $U_R^{(q)}$ that appear in the
decomposition of $U_R$ a number of times greater than or equal to
their dimension $d_q$, i.e. those for which
\begin{equation}\label{eq:specialproperty}
  d_q \equiv \text{dim}\,\mathcal{M}^{(q)}_R \leq \text{dim}\,\mathcal{N}^{(q)}_R  \,.
\end{equation}
We denote the set of $q$ that label such irreps by $Q_R$ and, in
what follows, we will be restricting our attention to only these
irreps. Also, for irreps $q\in Q_R$, choose an arbitrary subspace
$\bar{\mathcal{N}}^{(q)}_R \subseteq \mathcal{N}^{(q)}_R$ with
dimension $d_q$, i.e., with dimension equal to that of
$\mathcal{M}^{(q)}_R$.

We now define a new Hilbert space $\bar{\mathcal{H}}_R$ as
\begin{equation}
  \label{eq:RFdecompG}
  \bar{\mathcal{H}}_R = \bigoplus_{q \in Q_R} \mathcal{M}^{(q)}_R \otimes \bar{\mathcal{N}}^{(q)}_R \,,
\end{equation}
which is of dimension
\begin{equation} \label{eq:DR}
  D_R \equiv \sum_{q\in Q_R} d_q^2 \,.
\end{equation}
The state of $R$ that we will use for our quantum RF is
\begin{equation}
  \label{eq:Chiribella}
  \left\vert e\right\rangle =\sum_{q \in Q_R}\sqrt{\frac{d_q}{D_R}}
  \sum_{m=1}^{d_q}\left\vert q,m\right\rangle \otimes \left\vert \phi_{q,m}\right\rangle \,,
\end{equation}
where $\{|q,m\rangle\}$ is an arbitrary basis for
$\mathcal{M}^{(q)}$, and $\{|\phi_{q,m}\rangle\}$ an arbitrary basis
for  $\bar{\mathcal{N}}^{(q)}$.  Note that the orbit of $|e\rangle$
under $G$ has support in $\bar{\mathcal{H}}_R$.

The embedding $\bar{\mathcal{N}}^{(q)}_R \subseteq
\mathcal{N}^{(q)}_R$ provides a way of embedding $|e\rangle$ in the
original Hilbert space $\mathcal{H}_R$, and in addition $U_R$ acts
on the Hilbert space $\bar{\mathcal{H}}_R$ in the obvious way.  If
$Q_R$ contained all irreps of $G$, then $U_R$ would be the (left)
regular representation, and for a Lie group the Hilbert space
$\bar{\mathcal{H}}_R$ would be infinite dimensional.  If the quantum
RF is of bounded size, then a limited set of irreps appear in $Q_R$.

We note that, for the problem of optimally encoding a reference
frame relative to a maximum likelihood figure of merit, given a
general Hilbert space $\mathcal{H}_R$ the optimal states will not
have this precise form~\cite{Chi04,BRSreview,Chi04b,Chi06}.
However, such optimal states do take the form of
Eq.~\eqref{eq:Chiribella} when restricted to $\bar{\mathcal{H}}_R$.
These are the states of interest here.

The restriction to irreps having the special property of
Eq.~(\ref{eq:specialproperty}) is in fact critical for our analysis,
because only in this case can we define a useful \emph{right} action
of $G$ on the Hilbert space~\cite{KMP04}.  Consider the
representation $V_R$ of $G$ defined by its action on the covariant
set~\eqref{eq:CovariantRFStates} as
\begin{equation}
  \label{eq:RightAction}
  V_R(h)|g\rangle = |gh^{-1}\rangle \,, \qquad g,h\in G\,.
\end{equation}
To obtain an explicit form for this right action in terms of the decomposition of Eq.~\eqref{eq:RFdecompG}, we
make use of the fact that the state $|e\rangle$ is maximally entangled across the virtual tensor products
$\mathcal{M}^{(q)}_R \otimes \bar{\mathcal{N}}^{(q)}_R$.  Thus, we have for any transformation $U_R^{(q)}(h)$ on
a subsystem $\mathcal{M}^{(q)}_R$ the identity
\begin{align}
  U_R^{(q)}(h) \otimes I |e\rangle &= I \otimes U_R^{(q)}(h)^T |e\rangle \nonumber \\
  &= I \otimes U_R^{(q)}(h^{-1})^*|e\rangle \,,
\end{align}
where ${}^T$ denotes the transpose, ${}^*$ the complex conjugate, and we have made use of the fact that
$U_R^{(q)}$ is unitary. Given that the complex conjugate of a representation $U_R^{(q)}$ of $G$ is also a
representation of $G$ (called the conjugate representation and denoted by $U_R^{(q^*)}$), we can define a
representation $V_R$ by
\begin{equation}
  \label{eq:ConjugateIrreps}
  V_R(h) = \bigoplus_{q \in Q_R} I \otimes V^{(q^*)}_R(h) \,.
\end{equation}
In contrast to $U_R$, the representation $V_R$ acts on the subsystems $\bar{\mathcal{N}}^{(q)}_R$ irreducibly
according to the conjugate representation $q^*$, and leaves the subsystems $\mathcal{M}^{(q)}_R$ invariant.
Clearly, the two actions $U_R$ and $V_R$ commute.  Furthermore, it is easy to verify that $V_R$ satisfies
Eq.~(\ref{eq:RightAction}).

As the states of the reference frame are restricted to the Hilbert
space $\bar{\mathcal{H}}_R$, it is useful to consider our encoding
operation $\mathcal{E}$ of Eq.~(\ref{eq:Encoding}) with fiducial
state $|e\rangle$ of Eq.~(\ref{eq:Chiribella}) as a map from
$\mathcal{B}(\mathcal{H}_S)$ to $\mathcal{B}(\bar{\mathcal{H}}_R
\otimes \mathcal{H}_S)$.  For the remainder of this paper, we
consider the encoding map to be defined thus.

Finally, we note that the map $\mathcal{E}$ with the fiducial state $|e\rangle$ chosen to be of the
form~(\ref{eq:Chiribella}) is unital.  (Because the input and output spaces of $\mathcal{E}$ are of differing
dimension, we define \emph{unital} for such a trace-preserving map as one which maps the (normalized) completely
mixed state to the completely mixed state.) This result is seen as
\begin{align}
  \mathcal{E}(I_S/d_S) &= \int \text{d}g\, |g\rangle \langle g| \otimes
  I_S/d_S \nonumber \\
  &= \frac{1}{D_R} \sum_{q\in Q_R} I_{\mathcal{M}^{(q)}_R} \otimes
  I_{\bar{\mathcal{N}}^{(q)}_R} \otimes I_S/d_S \nonumber \\
  &= \frac{I_{\bar{\mathcal{H}}_R}}{D_R} \otimes \frac{I_S}{d_S} \,,
  \label{eq:unital}
\end{align}
where $d_S$ is the dimension of $\mathcal{H}_S$.  Here, we have used the fact that the maximally-entangled
states $\sum_{m=1}^{d_q} |q,m\rangle \otimes |\phi_{q,m}\rangle$ in Eq.~(\ref{eq:Chiribella}) have reduced
density matrices on ${\bar{\mathcal{N}}^{(q)}_R}$ that are proportional to the identity.

\subsection{Relational subsystems}
\label{subsec:RelSubsystems}

It is illustrative to investigate the action of the encoding map~\eqref{eq:Encoding} (the fiducial state
$|e\rangle$ will be assumed to be given by Eq.~(\ref{eq:Chiribella}) except in the final section of the article)
and to explicitly identify the subsystems of $\bar{\mathcal{H}}_{RS} \equiv \bar{\mathcal{H}}_R \otimes
\mathcal{H}_S$ into which the system's state is encoded.
The details of this section require extensive use of the virtual tensor product structure of
$\bar{\mathcal{H}}_{RS}$ induced by the unitary representation $U_{RS}$ of $G$, given in
Eq.~\eqref{eq:HdecompFull}, as well as an application of the Stinespring theorem for covariant maps~\cite{KW99};
this section may be skipped on first reading.  To facilitate this, we first state the main result of this
section prior to our detailed investigation of the encoding map.

\begin{description}
\item[Main result:]
According to Eq.~\eqref{eq:HdecompFull}, the joint Hilbert space $\bar{\mathcal{H}}_{RS}$ can be decomposed under the representation $U_{RS}$ of $G$ as
\begin{equation}
    \label{eq:HdecompRS}
  \bar{\mathcal{H}}_{RS} = \bigoplus_{q \in Q_{RS}} \mathcal{M}^{(q)}_{RS} \otimes \bar{\mathcal{N}}^{(q)}_{RS} \,,
\end{equation}
where $Q_{RS}$ are the set of irreps $q$ of $G$ that appear in the decomposition of $U_{RS}$. The encoding
map~\eqref{eq:Encoding} yields $G$-invariant density operators which, in terms of the
decomposition~\eqref{eq:HdecompRS}, are block-diagonal in the irreps $q\in Q_{RS}$ and, within each block, have
the form of a tensor product of the completely mixed state on the subsystem $\mathcal{M}^{(q)}_{RS}$ and some
non-trivial state on the subsystem $\bar{\mathcal{N}}^{(q)}_{RS}$.  Thus, the action of the encoding can be expressed
as
\begin{equation}
  \label{eq:Edecomp1}
  \mathcal{E}(\rho) =\sum_{q \in Q_{RS}} \bigl( d_q^{-1} I_{\mathcal{M}^{(q)}_{RS}}\bigr)
  \otimes \mathcal{E}^{(q)}(\rho) \,,
\end{equation}
where $I_{\mathcal{M}^{(q)}_{RS}}$ is the identity operator on
$\mathcal{M}^{(q)}_{RS}$, and $\mathcal{E}^{(q)}$ is a
trace-decreasing map from states on $\mathcal{H}_S$ to states on
$\bar{\mathcal{N}}^{(q)}_{RS}$. We show below that, under the
assumption that $\mathcal{H}_S$ is an irrep of $G$, each of these
encodings $\mathcal{E}^{(q)}$ takes the form
\begin{equation}
  \label{eq:Edecomp2}
  \mathcal{E}^{(q)}(\rho) = \frac{d_q}{D_R} A^{(q)\dag} \bigl( I_{\mathcal{K}^{(q)}} \otimes \rho\bigr) A^{(q)} \,,
\end{equation}
where $I_{\mathcal{K}^{(q)}}$ is the identity operator on a Hilbert space $\mathcal{K}^{(q)}$ carrying an irrep
$q^*$ of $G$, and $A^{(q)}:\bar{\mathcal{N}}^{(q)}_{RS} \rightarrow \mathcal{K}^{(q)} \otimes \mathcal{H}_S$ is
a linear map satisfying $A^{(q)\dag}A^{(q)} = I_{\bar{\mathcal{N}}^{(q)}_{RS}}$. In addition, each map $A^{(q)}$
takes a very simple form, which depends on the irrep $q$. Specifically, there is a subset of irreps $Q^{\rm
ok}_{RS} \subset Q_{RS}$ such that, for $q \in Q^{\rm ok}_{RS}$, the map $A^{(q)}$ is a bijective isometry, that
is,  a unitary; in these instances, the map $\mathcal{E}^{(q)}$ can be inverted and $\rho$ can be recovered
perfectly. For $q$ not in $Q^{\rm ok}_{RS}$, the map $A^{(q)}$ is an isometry that is not surjective, i.e., it
maps onto a \emph{proper subspace} of $\mathcal{K}^{(q)} \otimes \mathcal{H}_S$. The map $\mathcal{E}^{(q)}$ is
not invertible in these cases.
\end{description}

We can identify the relational degrees of freedom in which the
message is encoded by investigating how relational transformations
act on the Hilbert space $\bar{\mathcal{H}}_{RS}$.  The subsystems
$\mathcal{M}^{(q)}_{RS}$ carry an irreducible representation of $G$
corresponding to the collective action $U_{RS}$ and describe
collective degrees of freedom.  In contrast, the subsystems
$\bar{\mathcal{N}}^{(q)}_{RS}$ are relational.  However, not all
degrees of freedom in $\bar{\mathcal{N}}^{(q)}_{RS}$ describe
relations of the system $S$ to the reference frame $R$; some of
these describe relations among the parts of $R$ (or among the parts
of $S$ if the latter are composite systems).  We seek to identify,
for each irrep $q$, the precise subsystem of
$\bar{\mathcal{N}}^{(q)}_{RS}$ into which the message state $\rho$
is encoded.

The system Hilbert space $\mathcal{H}_S$ carries a representation $U_S$ of $G$.  If we act with $G$ on the
system but not on the RF, this will induce a relative transformation of the two; however, this action alone is
not $G$-invariant.  While it is possible to construct a $G$-invariant action of $U_S$ by using the techniques of
Refs.~\cite{KMP04,BRSreview}, it is much more straightforward to make use of the right action $V_R$ of $G$
defined in Eq.~(\ref{eq:RightAction}).  This action commutes with the left action $U_R$, and thus also commutes
with the collective action $U_{RS}$ of $G$.
By acting with $V_R(h)$ for $h\in G$ on a state $\rho_{RS} = \mathcal{E}(\rho)$ of the
form~\eqref{eq:GInvState}, we have
\begin{align}
  V_R(h) \mathcal{E}(\rho) V_R^\dag(h) &= \int \text{d}g\, |gh^{-1}\rangle \langle gh^{-1}| \otimes U_S(g)\rho U_S^{\dag }(g) \nonumber \\
  &= \int \text{d}g'\, |g'\rangle \langle g'| \otimes U_S(g'h)\rho U_S^\dag(g'h) \nonumber \\
  &= \mathcal{E}\bigl( U_S(h)\rho U_S^\dag (h) \bigr) \,,
  \label{eq:Gcovariant}
\end{align}
where we have used the invariance of the Haar measure.  The action of $V_R(h)$ on $\rho_{RS}$ yields another
invariant state, but one which is now an encoding of the transformed state $U_S(h)\rho U_S^\dag (h)$.  Thus,
$V_R(h)$ acts as a transformation of the relation between $S$ and $R$. A map $\mathcal{E}$ satisfying
Eq.~\eqref{eq:Gcovariant} is called \emph{$G$-covariant}.

As $V_R$ is a relational action, it acts on the subsystems
$\bar{\mathcal{N}}^{(q)}_{RS}$ in Eq.~\eqref{eq:HdecompRS}; we now
decompose these subsystems according to the irreps of $G$ under the
action of $V_R$, and in doing so identify the subsystems in which we
find the image of $\rho$ under the encoding map.

At this stage, we restrict our attention to the case where $U_S$ is
an irrep of $G$, labelled $q_S$. It appears straightforward
(although with substantially more burdensome notation) to extend our
results to the general case wherein this restriction is relaxed.
Indeed, the U(1) example considered in Sec.~\ref{sec:phase} provides
evidence of the generality of our theorem. However, we do not
consider the general case here.

Recall that the reference frame $R$ has a Hilbert space given by Eq.~\eqref{eq:RFdecompG}, and the system's
Hilbert space is $\mathcal{H}_S = \mathcal{M}^{(q_S)}_S$.  Thus,
\begin{align}
  \bar{\mathcal{H}}_{RS} &= \bar{\mathcal{H}}_R \otimes \mathcal{H}_S \nonumber \\
  &= \bigoplus_{q'\in Q_R} \Bigl( \mathcal{M}^{(q')}_R \otimes \mathcal{M}^{(q_S)}_S \Bigr)
  \otimes \bar{\mathcal{N}}^{(q')}_R  \nonumber \\
  &= \bigoplus_{q'\in Q_R} \Bigl( \bigoplus_{q |(q',q_S)\to q} \mathcal{M}^{(q)}_{RS}
  \otimes \mathcal{V}^{q',q_S}_{q} \Bigr) \otimes \bar{\mathcal{N}}^{(q')}_R \nonumber \\
  &= \bigoplus_{q\in Q_{RS}} \mathcal{M}^{(q)}_{RS} \otimes \Bigl( \bigoplus_{q'\in Q_R |(q',q_S)\to q}
  \bar{\mathcal{N}}^{(q')}_R \otimes \mathcal{V}^{q',q_S}_{q} \Bigr) \,,
  \label{eq:BigRSdecomp}
\end{align}
where $(q',q_S)\to q$ denotes that the irreps $q'$ and $q_S$ couple
to the irrep $q$, $\mathcal{V}^{q',q_S}_{q}$ is the multiplicity
space for the irrep $q$ in tensor representation $U^{(q')}_R \otimes
U_S$, and where $Q_{RS}$ is the set of all irreps that are obtained
by coupling some irrep $q' \in Q_R$ to $q_S$. Comparing the
expression above with Eq.~\eqref{eq:HdecompRS}, the subsystems
$\bar{\mathcal{N}}^{(q)}_{RS}$ are given by
\begin{equation}
  \label{eq:RecouplingNs}
  \bar{\mathcal{N}}^{(q)}_{RS} = \bigoplus_{q'\in Q_R|(q',q_S)\to q}
  \bar{\mathcal{N}}^{(q')}_{R} \otimes \mathcal{V}^{q',q_S}_{q} \,.
\end{equation}
We use the fact that if $(q',q_S)\to q$, then $(q^*,q_S)\to
(q')^*$~\cite{KW99}, and that $\mathcal{V}^{q',q_S}_{q} \simeq
\mathcal{V}^{q^*,q_S}_{(q')^*}$~\cite{KMP04}.  Thus,
\begin{equation}
  \bar{\mathcal{N}}^{(q)}_{RS} = \bigoplus_{ q'\in Q_R |(q^*,q_S)\to (q')^*}
  \bar{\mathcal{N}}^{(q')}_{R} \otimes \mathcal{V}^{q^*,q_S}_{(q')^*}  \,.
  \label{eq:RecouplingNsbar}
\end{equation}
where each subsystem $\bar{\mathcal{N}}^{(q')}_{R}$ on the
right-hand side carries an irrep $(q')^*$ of $G$ under the action
$V_R$.  (We note that in the examples presented in the latter
sections, with groups $U(1)$ and $SU(2)$, the subsystems
$\mathcal{V}$ are trivial and can be ignored.)

At this stage, we will make use of the $G$-covariance of the encoding map $\mathcal{E}$, given by
Eq.~\eqref{eq:Gcovariant}, to determine how the message is encoded into the relational subsystems
$\bar{\mathcal{N}}^{(q)}_{RS}$.  As the state $\mathcal{E}(\rho)$ is $G$-invariant under the action of $U_{RS}$ for
any $\rho$, it can be expressed according to the Hilbert space decomposition~\eqref{eq:HdecompRS} as
\begin{equation}
  \label{eq:Eofrho}
  \mathcal{E}(\rho) = \sum_{q \in Q_{RS}} \bigl( d_q^{-1} I_{\mathcal{M}^{(q)}_{RS}}\bigr) \otimes \mathcal{E}^{(q)}(\rho) \,,
\end{equation}
where $I_{\mathcal{M}^{(q)}_{RS}}$ is the identity operator on
$\mathcal{M}^{(q)}_{RS}$ and $\mathcal{E}^{(q)}(\rho)$ is an
(unnormalized) density operator on $\bar{\mathcal{N}}^{(q)}_{RS}$.
This expression defines a set of trace-decreasing superoperators
$\mathcal{E}^{(q)}: \mathcal{B}(\mathcal{H}_S) \rightarrow
\mathcal{B}(\bar{\mathcal{N}}^{(q)}_{RS})$; note that the latter can
be naturally embedded in the full multiplicity spaces
$\mathcal{N}^{(q)}_{RS}$ of the combined system.  From
Eq.~(\ref{eq:unital}), we see that the maps $\mathcal{E}^{(q)}$ are
also unital in that $\mathcal{E}^{(q)}(I_S)$ is proportional
(because $\mathcal{E}^{(q)}$ is trace-decreasing) to the identity on
$\bar{\mathcal{N}}^{(q)}_{RS}$. Also, as $V_R$ commutes with
$U_{RS}$, we have that each term $\mathcal{E}^{(q)}$ is itself
$G$-covariant, satisfying $V_R^{(q^*)}(h) \mathcal{E}^{(q)}(\rho)
V_R^{(q^*)\dag}(h) = \mathcal{E}^{(q)}( U_S(h)\rho U_S^\dag (h) )$.

We now make use of the Stinespring theorem for covariant CP
maps~\cite{Scu79}; in particular, we use a form due to Keyl and
Werner~\cite{KW99} for unital covariant CP maps.  There exists
another unitary representation $W^{(q^*)}$ of $G$ on a space
$\mathcal{K}^{(q)}$ and an \emph{intertwiner} (linear map)
\begin{equation}
  A^{(q)}: \bar{\mathcal{N}}^{(q)}_{RS} \rightarrow \mathcal{K}^{(q)} \otimes \mathcal{H}_S\,,
\end{equation}
satisfying $A^{(q)\dag}A^{(q)} = I_{\bar{\mathcal{N}}^{(q)}_{RS}}$ with
\begin{equation}
  \label{eq:Intertwiner}
  A^{(q)} V_R^{(q^*)}(h) = W^{(q^*)}(h) \otimes U_S (h) A^{(q)}\,,
\end{equation}
such that
\begin{equation}
  \label{eq:Eq}
  \mathcal{E}^{(q)}(\rho) = \frac{d_q}{D_R} A^{(q)\dag} \bigl(I_{\mathcal{K}^{(q)}} \otimes \rho \bigr)A^{(q)}\,.
\end{equation}

The form of Eq.~\eqref{eq:RecouplingNsbar} allows us to identify a
suitable Stinespring extension.  The representation $V_R^{(q^*)}$
acts on $\bar{\mathcal{N}}^{(q)}_{RS}$ through what appears to be
(ignoring the limits on the sum $q'\in Q_R$) a tensor representation
of an irrep $q^*$ with an irrep $q_S$. Thus, we can choose our
Stinespring extension in a minimal way such that $W^{(q^*)}$ acts on
$\mathcal{K}^{(q)}$ irreducibly as the irrep $q^*$ of $G$.  The
operators $A^{(q)}$ intertwine the representation $V_R^{(q^*)}$ with
the collective representation $W^{(q^*)}\otimes U_S$ on
$\mathcal{K}^{(q)} \otimes \mathcal{H}_S$. We now consider two
cases:

\textbf{Case A:}  If $q\in Q_{RS}$ is such that, for all $q'$
obtained via $(q^*, q_S) \rightarrow (q')^*$ then $q' \in Q_R$,
(i.e., $Q_R$ contains all of the irreps $q'$ that one can obtain by
$(q^*, q_S) \rightarrow (q')^*$), then the direct sum Hilbert space
in Eq.~\eqref{eq:RecouplingNsbar} is given by
\begin{equation}
  \label{eq:Refactorize}
  \bar{\mathcal{N}}^{(q)}_{RS} \simeq \mathcal{K}^{(q)} \otimes \mathcal{H}_S \,,
\end{equation}
where $\simeq$ denotes that these spaces are unitarily equivalent;
that is, the map $A^{(q)}$ is a bijective isometry and simply
represents the Clebsch-Gordan transformation relating the tensor
product of two irreps with the direct sum decomposition of
$\bar{\mathcal{N}}^{(q)}_{RS}$ given in~\eqref{eq:RecouplingNsbar}.
Let $Q_{RS}^{\rm ok} \subseteq Q_{RS}$ denote the set of irreps
satisfying this condition.

\textbf{Case B:}  If, however, $q\in Q_{RS}$ is such that the
condition of case A fails (i.e. $Q_R$ does \emph{not} contain all of
the irreps $q'$ that one can obtain by coupling $q^*$ and $q_S$),
then the intertwiner is no longer surjective.  Rather, the
intertwiner maps $\bar{\mathcal{N}}^{(q)}_{RS}$ onto a proper
subspace of $\mathcal{K}^{(q)} \otimes \mathcal{H}_S$, specifically
the subspace defined by the carrier space of the irreps $(q')^*$,
with $q' \in Q_R$, obtained through the coupling of the irrep $q^*$
on $\mathcal{K}^{(q)}$ with the irrep $q_S$ on $\mathcal{H}_S$.
(This space is necessarily a proper subspace of $\mathcal{K}^{(q)}
\otimes \mathcal{H}_S$ because, by the conditions of case B, $Q_R$
does not contain all irreps obtained in this coupling.)  A set of
basis states for this subspace can be calculated explicitly in any
particular instance using the Clebsch-Gordan coefficients for the
group $G$.

We now turn to the probabilities of each of these cases. We note
that $p_q = {\rm Tr}[\Pi_q \mathcal{E}(\rho)] ={\rm
Tr}[\mathcal{E}^{(q)}(\rho)]$ is the probability that the system is
encoded into the irrep $q$. We now prove that, for the case where
$\mathcal{H}_S$ carries an irrep $U_S$ of $G$, this probability is
independent of $\rho$. Using Eq.~(\ref{eq:Eq}), we have
\begin{align}
  p_q &= \frac{d_q}{D_R} {\rm Tr}_{\bar{\mathcal{N}}^{(q)}_{RS}}\bigl[ A^{(q)\dag} \bigl(I_{\mathcal{K}^{(q)}} \otimes \rho
  \bigr)A^{(q)} \bigr] \nonumber \\
  &= \int {\rm d}g \frac{d_q}{D_R} {\rm Tr}_{\bar{\mathcal{N}}^{(q)}_{RS}}\bigl[V^{(q^*)}(g) A^{(q)\dag} \bigl(I_{\mathcal{K}^{(q)}} \otimes \rho
  \bigr) \nonumber \\
  &\qquad \times A^{(q)} V^{(q^*)}(g)^{-1} \bigr] \nonumber \\
  &= \frac{d_q}{D_R} {\rm Tr}_{\bar{\mathcal{N}}^{(q)}_{RS}}\bigl[ A^{(q)\dag} \bigl(I_{\mathcal{K}^{(q)}} \otimes
  \int {\rm d}g\, U_S(g)\rho U_S(g)^{-1}
  \bigr)A^{(q)} \bigr], \nonumber \\
\end{align}
where in the second line we have used the G-invariance of $\mathcal{E}^{(q)}$ and in the third line we have used
Eq.~(\ref{eq:Intertwiner}). Because $\mathcal{H}_S$ is an irrep, it follows that $\int {\rm d}g\, U_S(g)\rho
U_S(g)^{-1} = I_S/d_S$ where $d_S = {\rm dim}\mathcal{H}_S$, and therefore $p_q$ is independent of $\rho$ for
all $q$.  The SU(2) case, presented in Sec.~\ref{sec:Cartesian}, provides an explicit example of this.

We note that for the general case, where the system does not carry
an irrep of $G$, then this probability can be state-dependent.  For
example, if $\mathcal{H}_S$ is a direct sum of irreps $q_S$, then
$\int {\rm d}g\, U_S(g)\rho U_S(g)^{-1} = \sum_{q_S}
\textrm{Tr}(\rho\Pi^{(q_S)}_S)\Pi^{(q_S)}_S$ where $\Pi^{(q_S)}_S$
is the projector onto the $q_S$ irrep of $\mathcal{H}_S$.  In this
case, for $q \in Q_{RS}^{\textrm{ok}}$ (where $A^{q}$ is unitary),
we have $p_q =d_q^2/D_R$, independent of $\rho$.  However, for $q
\notin Q_{RS}^{\textrm{ok}}$, the weight $p_q$ can depend on $\rho$.
This occurs for the U(1) case, as seen explicitly in
Sec.~\ref{sec:phase}.

With each map $\mathcal{E}^{(q)}$ now defined through Eq.~(\ref{eq:Eq}), we can explicitly express $\mathcal{E}$
in Kraus operator form as $\mathcal{E}(\rho) = \sum_{q,m,\mu} K_{q,m,\mu} \rho K^\dag_{q,m,\mu}$,
where
\begin{equation}
  \label{eq:KrausE}
  K_{q,m,\mu} = \frac{1}{\sqrt{D_R}} |q,m\rangle \otimes A^{(q)\dag}|q,\mu\rangle\,,
\end{equation}
and where $|q,m\rangle$ is a basis for $\mathcal{M}^{(q)}_{RS}$ and $|q,\mu\rangle$ is a basis for $\mathcal{K}^{(q)}$.

Finally, we point out a useful expression for $\mathcal{E}^{(q)}$
(which applies regardless of whether $\mathcal{H}_S$ carries an
irrep of $G$).  From Eq.~(\ref{eq:Eofrho}), it is clear that
\begin{equation}
  \mathcal{E}^{(q)}(\rho)=\textrm{Tr}_{\mathcal{M}_{RS}^{(q)}} \bigl( \Pi_q \mathcal{E}(\rho) \Pi_q \bigr) \,.
\end{equation}
Combining this with the expression for $\mathcal{E}$ in Eq.~(\ref{eq:GInvState}) and making use of
Eq.~(\ref{eq:GeneralDecoherenceDfullDfree}), we obtain
\begin{equation} \label{eq:Equsefulform}
\mathcal{E}^{(q)}(\rho)=\textrm{Tr}_{\mathcal{M}_{RS}^{(q)}} \bigl[ \Pi_q (|e\rangle \langle e| \otimes \rho)
 \Pi_q \bigr]\,.
\end{equation}
This form will be used frequently when working out explicit
examples.

\section{Decoding}
\label{sec:decoding}

In the communication protocol we are considering, Bob's task is to
recover the quantum message $\rho$ by implementing a decoding map
$\mathcal{R}: \mathcal{B}(\bar{\mathcal{H}}_{R}\otimes
\mathcal{H}_{S}) \rightarrow \mathcal{B}(\mathcal{H}_S)$.  A useful
recovery map to consider is the following
\begin{equation}
  \label{eq:decoding}
  \mathcal{R} = D_R \mathcal{E}^\dag \,,
\end{equation}
where the adjoint for superoperators is defined relative to the Hilbert-Schmidt inner product, $\mathrm{Tr}(A
\mathcal{E}^{\dag}[B])= \mathrm{Tr}(\mathcal{E}[A] B)$.  The map $\mathcal{R}$ is completely positive and linear, because the superoperator adjoint preserves these features.  It
is also trace-preserving.  To see this fact, observe that $\mathcal{E}(I_S)= (I_R/D_R) \otimes I_S$, where $I_R= D_R \int {\rm d}g |g\rangle \langle g|$ is the identity operator on $\bar{\mathcal{H}}_R$; consequently, for $\rho_{RS} \in \mathcal{B}(\bar{\mathcal{H}}_{R}\otimes \mathcal{H}_{S})$, we have $\mathrm{Tr}[\mathcal{R}(\rho_{RS})]=\mathrm{Tr}[\rho_{RS}\mathcal{R}^{\dag}(I_{S})] = \mathrm{Tr}[\rho_{RS}]$. We have therefore
verified that $\mathcal{R}$ is a valid quantum operation that can be implemented deterministically.

Assuming that there is no prior information about the input state to $\mathcal{E}$, the map $D_R
\mathcal{E}^{\dag}$ is precisely the ``approximate reversal'' operation for $\mathcal{E}$ proposed by Barnum and
Knill~\cite{Bar02} which yields an error no more than twice that of the optimal reversal operation. The error
here is defined in terms of the deviation from unity of the average entanglement fidelity for an arbitrary input
ensemble.

\subsection{``Measure and re-orient'' implementation of decoding}

Given a superoperator $\mathcal{A}$, the adjoint $\mathcal{A}^\dag$ is easily determined through a Kraus
decomposition of $\mathcal{A}$.  Specifically, if $\mathcal{A}[\rho]=\sum_{i} K_{i}\rho K_{i}^{\dag}$ then
$\mathcal{A}^{\dag}[\rho]=\sum_{i}K_{i}^{\dag }\rho K_{i}$.  The expression~(\ref{eq:Encoding}) for the encoding
map $\mathcal{E}$ provides one Kraus decomposition: the covariant set of operators $\{K(g), g\in G\}$, where
$K(g)=|g\rangle \otimes U_{S}(g)$.  It follows that a covariant set of Kraus operators for $\mathcal{E}^{\dag}$
is $\{K^{\dag}(g), g\in G\}$ where $K^{\dag}(g)=\langle g| \otimes U_{S}^{\dag}(g)$ and consequently
\begin{equation}
  \mathcal{R}[\rho_{RS}] = D_R \int\mathrm{d}g\bigl(  \langle g| \otimes U_{S}^{\dag}(g)\bigr) \rho_{RS}\bigl( |g\rangle \otimes U_{S}(g)\bigr)  \,.
\end{equation}
From this expression, we see that one way in which $\mathcal{R}$ can
be implemented is as follows: measure the covariant POVM $\{ D_R
|g\rangle\langle g| \mathrm{d}g\}$ on $R$, then implement the
unitary $U_{S}(g^{-1})$ on $S$, and finally discard $R$ and the
measurement result $g$.  We refer to this as the ``measure and
re-orient'' implementation of the decoding map.  (It is the adjoint
of the ``prepare and $G$-twirl'' implementation of the encoding
map.)

So we see that the decoding map we are considering is in fact the most obvious recovery scheme one can imagine!
Bob simply estimates the relative orientation between Alice's reference frame and his own by measuring how the
sample $R$ of Alice's RF is oriented, then re-orients the system appropriately (i.e. in such a way that it is
finally oriented relative to his RF in precisely the way that it was initially oriented relative to Alice's RF).
One can view the system $R$ as a cryptographic key or calibrating system that contains the information for how
to recover the quantum state of $S$.  It is noteworthy that this implementation of Bob's decoding map does not
require any entangling operations between $R$ and $S$.  Bob can achieve it with local operations and classical
communication (LOCC) between $R$ and $S$.  Because Bob does not need to possess $S$ to implement the
appropriate measurement on $R$, it follows that Alice can subsequently transmit an arbitrary number of systems
and Bob can decode these with the same fidelity as the first.

\subsubsection*{Effective decoherence}

Consider the action of the decoding map on states $\rho_{RS} = \mathcal{E}(\rho)$, i.e., on states of the form
of Eq.~\eqref{eq:Encoding}.  After a measurement on $R$ having outcome $g'$, followed by a transformation
$U_S(g')^{-1}$ to system $S$, the reduced density operator on $S$ is
\begin{align}
  \mathcal{R} \circ \mathcal{E}[\rho] &= D_R \int \text{d}g\, |\langle g|g'\rangle |^{2}\, U_S((g')^{-1}g)\rho U_S^{\dag}((g')^{-1}g) \nonumber \\
  &= D_R \int \text{d}g\, | \langle e|g\rangle |^{2}\, U_S(g)\rho U_S^{\dag }(g)\,, \label{eq:Method2}
\end{align}
where the simplification occurs because d$g$ is invariant.  Note
that the result is independent of the outcome $g'$. It is
straightforward to check that this state is normalized, as $D_R^{-1}
= \int \text{d}g\, |\langle e|g\rangle |^{2}$.  This is precisely
how a state $\rho$ relative to Alice's frame would be redescribed
relative to Bob's frame if their relative orientation $g$ was known
to be distributed according to the probability distribution $p(g) =
D_R | \langle e|g\rangle |^{2}$.  If $|\langle g|e\rangle|^{2}$ as a
function of $g$ is highly peaked around the identity group element
$e,$ then the only unitary that will contribute significantly in the
integral will be the identity operation, and we will have
$\mathcal{R}\circ\mathcal{E}[\rho]\simeq\rho$.  It is the narrowness
of the distribution $|\langle g|e\rangle|^{2}$, a measure of the
quality of the quantum reference frame, that determines the degree
to which one can recover the quantum information.

We see that for bounded-size samples of Alice's reference frame, the decoding map we have described achieves
\emph{approximate error correction}.  Further on, we will show that the degree to which it deviates from perfect
error correction is inversely proportional to the size of the quantum reference frame.

\subsection{``Extract from the relational subsystems'' implementation of decoding}

Recall that the ``measure and re-orient'' implementation of the recovery operation $\mathcal{R}$ was inferred
from the adjoint of the Kraus decomposition $\{|g\rangle \otimes U_S(g)\, | \,g\in G\}$  of $\mathcal{E}$.   We
exhibited a different Kraus decomposition of the encoding operation in Eq.~(\ref{eq:KrausE}).
The adjoint of the latter provides a novel Kraus decomposition of $\mathcal{R}$ and therefore also a new way of
implementing the recovery operation.  We will refer to it as the ``extract from the relational subsystems''
implementation.  We find that $\mathcal{R} = D_R \mathcal{E}^\dag$ can be written as
\begin{equation} \label{eq:decodingdecomposition}
  \mathcal{R}(\rho_{RS})=\sum_{q\in Q_{RS}} \mathcal{R}^{(q)}\bigl({\rm Tr}_{\mathcal{M}^{(q)}_{RS}} \bigl[\Pi^{(q)} \rho_{RS} \Pi^{(q)}\bigr]\bigr) \,,
\end{equation}
where we define
\begin{equation} \label{eq:Rq}
  \mathcal{R}^{(q)}(\cdot) = \mathrm{Tr}_{\mathcal{K}^{(q)}} \bigl[  A^{(q)}(\cdot)A^{(q)\dag}\bigr] \,,
\end{equation}
as a map from $\mathcal{B}(\bar{\mathcal{N}}^{(q)}_{RS})$ to
$\mathcal{B}(\mathcal{H}_S)$. Recalling the form of the encoding map
$\mathcal{E}^{(q)}$ of Eq.~(\ref{eq:Edecomp2}), we see that
$\mathcal{R}^{(q)} = d_q \mathcal{E}^{(q)\dag}$.

This implementation of the decoding map $\mathcal{R}$ differs from
the ``measure and re-orient'' scheme in that it requires joint (i.e.
nonseparable) operations on $R$ and $S$.  Specifically, it is
implemented via a joint unitary on $RS$ followed by a trace on $R$.

Finally, we highlight another decomposition of $\mathcal{R}^{(q)}$
that will be useful to us further on. It is the one obtained by
taking the adjoint of Eq.~(\ref{eq:Equsefulform}),
\begin{equation} \label{eq:Rqusefulform}
\mathcal{R}^{(q)}(\cdot)=d_q \langle e| I_{\mathcal{M}_{RS}^{(q)}}\otimes(\cdot)|e\rangle\,.
\end{equation}


\subsubsection*{Effective decoherence}

Given Eqs.~(\ref{eq:Edecomp1}) and (\ref{eq:decodingdecomposition}),
the composition $\mathcal{R}\circ\mathcal{E}$ can be written as
\begin{equation}
  \mathcal{R}\circ\mathcal{E}[\rho] = \sum_{q\in Q_{RS}} \mathcal{R}^{(q)}\circ\mathcal{E}^{(q)}[\rho]\,.
 \label{eq:composition}
\end{equation}
Substituting the expressions for $\mathcal{E}^{(q)}$ and $\mathcal{R}^{(q)}$ in Eqs.~(\ref{eq:Eq}) and
(\ref{eq:Rq}), we obtain
\begin{equation}
  \mathcal{R}\circ\mathcal{E}[\rho] = \sum_{q\in Q_{RS}} \frac{d_q}{D_R} \mathrm{Tr}_{\mathcal{K}^{(q)}}\bigl[
  A^{(q)}A^{(q)\dag}\bigl( I_{\mathcal{K}^{(q)}}\otimes\rho\bigr) A^{(q)}A^{(q)\dag}\bigr]\,.
 \label{eq:composition2}
\end{equation}

We now consider the two subsets of $Q_{RS}$ from
Sec.~\ref{subsec:RelSubsystems}.  In Case A, where $q\in
Q_{RS}^{\mathrm{ok}}$, the intertwiner $A^{(q)}$ is a bijective
isometry, and consequently $A^{(q)}A^{(q)\dag}$ is the identity and
$\mathcal{R}^{(q)}\circ\mathcal{E}^{(q)}[\rho] = (d_q^2/D_R) \rho$.
Therefore, in this case the quantum information is perfectly
recovered by the decoding map.  In Case B, however, $P^{(q)} =
A^{(q)}A^{(q)\dag}$ is a non-trivial projection on
$\mathcal{K}^{(q)} \otimes \mathcal{H}_S$ and the recovery is not
perfect.  We can express Eq.~(\ref{eq:composition2}) as
\begin{multline}
  \mathcal{R}\circ\mathcal{E}[\rho]
 = \Bigl( \sum_{q\in Q_{RS}^{\mathrm{ok}}} \frac{d_q^2}{D_R} \Bigr) \rho \\
 +  \sum_{q\notin Q_{RS}^{\mathrm{ok}}} \frac{d_q}{D_R} \mathrm{Tr}_{\mathcal{K}^{(q)}}\bigl[P^{(q)}\bigl( I_{\mathcal{K}^{(q)}}\otimes\rho\bigr) P^{(q)}\bigr]\,.
 \label{eq:composition3}
\end{multline}
This is just an alternative Kraus decomposition of the effective decoherence map of Eq.~(\ref{eq:Method2}).

\subsection{Comparison of implementations} \label{sec:Comparison}

We have shown two very distinct ways of implementing one and the
same decoding operation. If we describe the reference frame token
$R$ as an ancilla, then what we have is an example wherein a single
map can be implemented either by a joint unitary followed by a trace
on the ancilla, or by a measurement of the ancilla followed by a
unitary rotation on the system that depends on the outcome of the
measurement. The existence of many different implementations of an
operation is familiar in quantum information theory.  For instance,
Griffiths and Niu have made use of a similar multiplicity of
possibilities for the optimal eavesdropping strategies in quantum
cryptography~\cite{Niu99}.


The multiplicity of ways of implementing a single operation is analogous to the multiplicity of mixtures that
lead to the same density operator. Two Kraus decompositions of our $G$-invariant recovery operation differ in
their transformation properties under the group: one is a $G$-covariant set of operators (a continuous set in
the case of a Lie group) and the other is a discrete set of $G$-invariant operators.  Similarly, a $G$-invariant
density operator $\rho$ on a finite-dimensional Hilbert space admits two sorts of convex decompositions: a
spectral decomposition with a discrete number of $G$-invariant elements, and the decomposition induced by
$\rho$-distortion of a G-covariant POVM (continuous if $G$ is a Lie group) \cite{HJW93}.

Recognizing this multiplicity of convex decompositions and the fact that no particular decomposition is
preferred has been important for resolving many conceptual confusions~\cite{BRS06}. Furthermore, each
decomposition may yield important insights.  In quantum optics, for example, a Poissonian mixture of number
eigenstates is equivalent to a uniform mixture over coherent states with the same mean number but differing in
phase.  The decomposition into states with well-defined phase is particularly useful for making predictions
about wave-like phenomena, such as interference experiments, whereas the number state decomposition is best for
particle-like phenomena, such as determining number statistics~\cite{SBRK03}.


Similarly, each of the two decompositions we have provided of our
decoding operation provides some insight into our problem.  The
``measure and re-orient'' scheme is clearly the most intuitive and
demonstrates that joint operations on reference token and system are
not necessary to implement our recovery map. The ``extract from
relational subsystems'' scheme
demonstrates that if Bob begins by measuring the irrep of the
composite of $R$ and $S$, he learns whether the state was in a
``good'' irrep or not and consequently whether or not he has
achieved a perfect decoding.  This sort of post-selectively perfect
decoding operation is discussed in the following section.


\subsection{Post-selectively perfect decoding}
\label{subsec:PostSelective}

Thus far we have only judged decoding schemes by their average
performance.  It is also possible to say something about the best
and worst case performance. The ``measure and re-orient'' scheme is
not particularly interesting in this regard: one achieves precisely
the same fidelity of recovery regardless of the outcome of the
covariant measurement on the RF token, so that the best and worst
case recoveries are equivalent to the average. On the other hand, in
the ``extract from the relational subsystems'' scheme, we found that
the fidelity of the recovery depends on the irrep of the composite
of RF token and system into which the input state was encoded.
Furthermore, given that the decoding operation was incoherent over
these irreps, it is always possible to make a projective measurement
that distinguishes these. Depending on the measurement outcome, one
can achieve decodings with fidelities that are sometimes better and
sometimes worse than the average.

Indeed, by enhancing the ``extract from the relational subsystems''
scheme with such a measurement, Bob can achieve \emph{perfect}
decoding with some probability.  Specifically, if he finds one of
the ``good'' irreps, $q \in Q_{RS}^{\textrm{ok}}$, then
$\mathcal{E}^{(q)}$ is invertible and the decoding operation
$\mathcal{R}^{(q)}$ of Eq.~(\ref{eq:Rq}) recovers the quantum
message perfectly. (Of course, if he achieves one of the ``bad''
irreps, $q \notin Q_{RS}^{\textrm{ok}}$, then he achieves a decoding
that is worse than the average.) Recalling
Eq.~(\ref{eq:composition3}) and making use of Eq.~(\ref{eq:DR}), the
probability of perfect recovery is
\begin{equation} \label{eq:pperfect}
p_{\textrm{perfect}} = \frac{1}{D_R} \sum_{q\in Q^{\textrm{ok}}_{RS}} d_q^2 =\frac{\sum_{q\in
Q^{\textrm{ok}}_{RS}} d_q^2}{\sum_{q'\in Q_{R}} d_{q'}^2}\,.
\end{equation}
Such a decoding scheme achieves post-selectively perfect error correction \cite{PB08}.  It is akin to achieving
unambiguous discrimination of a set of nonorthogonal quantum states~\cite{UD}.

For this implementation of the decoding, note that Bob must be able
to store the quantum token of Alice's reference frame coherently
until the time when he receives the message systems. Another point
worth noting: if Alice is sending a large number of systems and Bob
wishes to implement probabilistically perfect error correction on
some subset of them, he must wait until he has collected all of the
systems in that subset. The reason is that he must perform a joint
measurement on the composite of these and the RF token. Furthermore,
after his measurement is complete, he has disrupted the state of the
RF token and he can no longer achieve perfect error correction for
any other systems. The tradeoffs involved in such postselectively
perfect decoding schemes are an interesting topic for future
research.

Finally, we can consider modifying the encoding operation rather than the decoding in a similar way.  Note that
if, immediately after implementing her encoding operation, Alice implements a projective measurement of the
irrep of the composite of RF token and system, she can come to know whether a subsequent decoding operation will
achieve a perfect recovery or not. In addition, if it happens that Alice has a classical description of the
quantum message rather than merely having a sample, then she can prepare the quantum state many times and only
initiate transmission when her measurement finds one of the ``good'' irreps. In this case, the quantum message
is perfectly encoded into a pure $G$-invariant state of the composite of system and RF token. Such a scheme
therefore achieves a relational encoding akin to the one presented in Bartlett \emph{et al.} \cite{BRS03}. The
precise connection of our results to the latter encoding is an interesting topic for future research (as is the
application of the mathematical tools developed here to the general problem of calibration-free communication
schemes discussed in the introduction).


\section{Example:  Phase reference}
\label{sec:phase}

The quantum state of a harmonic oscillator is always referred to
some phase reference~\cite{BRS06}.  In this example, we consider
using one quantum harmonic oscillator (a single mode) as a phase
reference for another, and investigate the effect of bounding the
maximum number $N_R$ of excitations in the phase reference.
Specifically, consider the single-mode RF to be prepared in the
bounded-size phase eigenstate
\begin{equation}
  \left\vert e_{N_R}\right\rangle
  =\frac{1}{\sqrt{N_{R}+1}}\sum_{n=0}^{N_{R}}\left \vert n\right\rangle\,,
\end{equation}
where $|n\rangle$ is the Fock state with $n$ excitations.  This
state is of the form of our general state (\ref{eq:Chiribella}) for
the case of $G=$ U(1).  For the system, we consider a qubit encoded
in the two-dimensional subspace spanned by $|0\rangle$ and
$|1\rangle$.  Note that the system we consider does not carry an
irrep of U(1), and in fact U(1) has only one-dimensional irreps.
Because our main result concerning the representation of the
encoding map, presented in Sec.~\ref{subsec:RelSubsystems}, was only
proven under the assumption that $\mathcal{H}_S$ is an irrep, the
U(1) example cannot be presented as a special case of this result.
Nonetheless, we find the U(1) example to be in accord with the
general result, suggesting that our theorem applies more generally.

For simplicity, we consider a system prepared in an arbitrary pure state
\begin{equation}
  \rho = |\psi\rangle \langle \psi|\,, \qquad |\psi\rangle = \alpha \left\vert 0\right\rangle +\beta \left\vert 1\right\rangle \,.
\end{equation}
Our results will directly extend to the mixed-state case via the linearity of convex combination.

\subsection{Effective decoherence}

The overlap of the RF state $|e_{N_R}\rangle$ with its rotated version is
\begin{equation}
  |\langle e_{N_R}|U_R(\theta)|e_{N_R}\rangle |^{2}
  = \Bigl| \sum_{n=0}^{N_{R}}e^{i\theta n} \Bigr|^{2}
  =\tfrac{1-\cos \left( N_{R}+1\right) \theta }{1-\cos \theta }\,.
\end{equation}
Rotations in U(1) act on the qubit system state as
\begin{equation}
  U_S(\theta )\rho U_S(\theta)^{\dag } =
  \left(\begin{smallmatrix}
  |\alpha |^{2} & \alpha \beta ^{\ast }e^{i\theta } \\
  \alpha ^{\ast }\beta e^{-i\theta } & |\beta |^{2}
  \end{smallmatrix}\right)\,.
\end{equation}
Evaluating Eq.~(\ref{eq:Method2}) then gives
\begin{equation}
  \mathcal{R}\circ \mathcal{E}(\rho)
  \propto \int \tfrac{d\theta }{2\pi }\tfrac{1-\cos [( N_{R}+1) \theta]}{1-\cos \theta }
  \left(\begin{smallmatrix}
  |\alpha |^{2} & \alpha \beta ^{\ast }e^{i\theta } \\
  \alpha ^{\ast }\beta e^{-i\theta } & |\beta |^{2}
  \end{smallmatrix}\right)\,.
\end{equation}
Noting that
\begin{align}
  \int \tfrac{d\theta }{2\pi }\tfrac{1-\cos [( N_{R}+1) \theta]}{1-\cos \theta } &= N_{R}+1\,, \\
  \int \tfrac{d\theta }{2\pi }\tfrac{1-\cos [( N_{R}+1) \theta]}{ 1-\cos \theta }e^{i\theta } &= N_{R}\,,
\end{align}
which also gives the normalization, we have
\begin{align} \label{eq:partialdephasing}
  \mathcal{R} \circ \mathcal{E}(\rho)
  &=\tfrac{N_{R}}{N_{R}+1}
  \left(\begin{smallmatrix}
  |\alpha |^{2} & \alpha \beta ^{\ast } \\
  \alpha ^{\ast }\beta  & |\beta |^{2}
  \end{smallmatrix}\right) +\tfrac{1}{N_{R}+1}\left(\begin{smallmatrix}
  |\alpha |^{2} & 0 \\
  0 & |\beta |^{2}
  \end{smallmatrix}\right) \nonumber \\
  &= \Bigl( \tfrac{N_{R}}{N_{R}+1} \mathcal{I} + \tfrac{1}{N_{R}+1} \mathcal{G} \Bigr)[\rho] \, ,
\end{align}
where $\mathcal{G}$ here denotes the U(1)-twirling operation (the dephasing map).  It follows that in the
``measure and re-orient'' scheme for decoding, regardless of the outcome of the measurement, the reduced density
operator is with probability $N_{R}/(N_{R}+1)$ the state $\alpha |0\rangle +\beta |1\rangle$, while with
probability $1/(N_{R}+1)$ it is completely dephased in the $|0\rangle, |1\rangle$ basis.  The overall effect of
encoding and decoding is to implement a partial dephasing.

\subsection{Relational subsystems}

Because the irreps of U(1) are all one-dimensional, we have
$\textrm{dim}\,\mathcal{M}_{RS}^{(N)} =1$ and consequently, by
Eq.~(\ref{eq:Edecomp1}), the encoding operation $\mathcal{E}$ may be
expressed simply as $\mathcal{E}(\rho)=\sum_N \mathcal{E}^{(N)}
(\rho)$. By virtue of Eq.~(\ref{eq:Equsefulform}), each operation
$\mathcal{E}^{(N)} (\rho)$ may in turn be expressed as
\begin{equation}
  \mathcal{E}^{(N)} (\rho) = \Pi^{(N)} [|e_{N_R}\rangle\langle e_{N_R}| \otimes \rho]\Pi^{(N)} \,,
\end{equation}
which evaluates for different values of $N$ as:
\begin{equation}
  \mathcal{E}^{(N)} (\rho) =
  \begin{cases}
  \alpha\left\vert 0,0\right\rangle, &N=0 \\
  \alpha \left\vert N,0\right\rangle +\beta \left\vert
  N{-}1,1\right\rangle, &0{<}N{<}N_{R}{+}1 \\
  \beta\left\vert N_{R},1\right\rangle, &N=N_{R}+1\,.
  \end{cases}
\end{equation}
The decoding operation has the form $\mathcal{R}=\sum_N \mathcal{R}^{(N)}$ where $\mathcal{R}^{(N)}\propto
\mathcal{E}^{(N)\dag}$.  One easily verifies that $\mathcal{R}^{(N)}$ maps $|N,0\rangle$ to $|0\rangle$ and
$|N{-}1,1\rangle$ to $|1\rangle$, so that
\begin{equation}
\label{eq:U(1)encodingcases}
  (\mathcal{R}^{(N)} \circ \mathcal{E}^{(N)})[\rho] \propto
  \begin{cases}
  \left\vert 0\right\rangle, &N=0 \\
  \alpha \left\vert 0\right\rangle +\beta \left\vert 1\right\rangle,
  &0<N<N_{R}+1 \\
  \left\vert 1\right\rangle, &N=N_{R}+1\,.
  \end{cases}
\end{equation}
The probability of the outcome $N=0$ is $|\alpha|^{2}/(N_{R}+1)$, of $N=N_R+1$ is $|\beta|^{2}/(N_{R}+1)$, and
of each of the other outcomes is $1/(N_{R}+1)$.  Weighting the decoded states $\mathcal{R}^{(N)} \circ
\mathcal{E}^{(N)}(\rho)$ by these probabilities, we can verify that Eq.~(\ref{eq:partialdephasing}) is
recovered.

In this particular example, taking the adjoint of the encoding
operation as one's recovery operation is actually optimal. The proof
is as follows. For $0<N<N_{R}+1$, the recovery operation is perfect
and consequently optimal. Otherwise, the action of the encoding map
is to measure the system in the $\{|0\rangle,|1\rangle\}$ basis and
update it to one of two orthogonal states [as can be inferred from
Eq.~(\ref{eq:U(1)encodingcases})]. It is a well-known result that
the update map that maximizes the entanglement fidelity is simply
the L\"{u}ders rule (or projection postulate)~\cite{Bar02b}, and
this is precisely what the composition of the encoding with the
recovery operation achieves.

The fact that the optimal recovery operation can be achieved using a
``measure and reorient'' scheme shows that having the classical
resource of partially correlated reference frames that is obtained
by this scheme is just as good as having the quantum reference frame
token, at least for the purpose of optimizing average-case
performance in decoding. This is a surprising result because one
might have expected the quantum resource to always do better.

Finally, by implementing a projective measurement of the total
number and post-selecting on finding $N \ne 0,N_R+1$, it is clear
that Bob can achieve perfect decoding.  This occurs with probability
\begin{equation}
  p_{\textrm{perfect}}=\frac{N_R}{N_R +1}\,.
\end{equation}

\section{Example:  Cartesian frame}
\label{sec:Cartesian}

For a Cartesian frame, the relevant group is the rotation
group.\footnote{We use $SU(2)$ rather than $SO(3)$ to allow for
spinor representations of the rotation group.}  The charge sectors
(irreps) are labeled by a nonnegative integer or half-integer $j$,
and the irreps are $(2j+1)$-dimensional with the standard basis
$\{|j,m\rangle, m=-j,\ldots,j\}$.  We bound the size of our
reference frame token by bounding $j$.  Recall that the fiducial
state for the frame, Eq.~(\ref{eq:Chiribella}), requires us to work
in a subspace $\mathcal{H}'_R \subseteq \mathcal{H}_R$ satisfying
Eq.~(\ref{eq:specialproperty}).  In the Cartesian case, we are
confined to $j$ values such that $\textrm{dim}\,\mathcal{N}_R^{(j)}
\ge 2j+1$.  We denote the largest such value by $j_R$. (As an
example, for an even number $N$ of spin-1/2 particles, only the
highest irrep, $j=N/2$, fails to satisfy
Eq.~(\ref{eq:specialproperty}), and consequently $j_R=N/2-1$.
See~\cite{Chi04,BRSreview}.)  For simplicity, we restrict our
attention to integer values of $j_R$ (similar results can be
obtained if one also allows non-integer values).  The fiducial state
of the RF token is
\begin{equation}
  \label{eq:SpinChiribella}
  |e_{j_R} \rangle =\sum_{j=0}^{j_R} \sqrt{\frac{2j+1}{D_R}} \sum_{m=-j}^j | j,m \rangle \otimes | \phi_{j,m}\rangle \,,
\end{equation}
where
\begin{equation} \label{eq:DRCartesian}
  D_R = \sum_{j=0}^{j_R} (2j+1)^2 = \frac{1}{3}(2j_R+1)(2j_R+3)(j_R+1).
\end{equation}

The system is taken to be a spin-$\frac{1}{2}$ particle. Because
this is an irrep of SU(2), the general results of
Sec.~\ref{subsec:RelSubsystems} apply.

\subsection{Effective decoherence}

We choose the following parametrization of SU(2),
\begin{equation}
  U(\omega,\theta,\phi) = e^{i \omega \mathbf{n}\cdot\mathbf{J}}\,,
\end{equation}
describing a rotation by angle $\omega$ about the unit vector $\mathbf{n} = (\sin\theta \cos\phi,\sin\theta
\sin\phi,\cos\theta)$. Let
\begin{equation}
  |(\omega,\theta,\phi)_{j_R} \rangle =U_R(\omega,\theta,\phi )\left\vert e_{j_R}\right\rangle\,,
\end{equation}
where $|e_{j_R}\rangle$ is the fiducial RF state of Eq.~(\ref{eq:SpinChiribella}).  Then
\begin{align}
  \langle e_{j_R}|(\omega,\theta,\phi)_{j_R}\rangle
  &=\sum_{j}\sum_{m}\frac{2j+1}{D_R}
  \left\langle j,m\right\vert U^{(j)}(\omega,\theta,\phi)\left\vert j,m\right\rangle \\
  &=\frac{1}{D_R}\sum_{j}(2j+1)\chi^{(j)}(\omega,\theta,\phi)\,,
\end{align}
where $U^{(j)}$ is the spin-$j$ irrep of SU(2), and
$\chi^{(j)}(\omega,\theta,\phi) =
\text{Tr}[U^{(j)}(\omega,\theta,\phi)]$ are the characters of SU(2).
These characters are independent of $\theta$ and $\phi$.  They are
given by
\begin{equation}
  \chi^{(j)}(\omega )=\frac{\sin [(j+\tfrac{1}{2})\omega]}{\sin[\omega /2]}\,,
\end{equation}
Using the following identity
\begin{equation}
  \frac{\sin[(n+1/2)\omega]}{2\sin[\omega/2]} = 1/2 + \sum_{k=1}^n \cos(k\omega)\,,
\end{equation}
we find that
\begin{multline}
  |\langle e_{j_R}|(\omega,\theta,\phi)_{j_R}\rangle |^{2} =
  \Bigl(\frac{\sin[\omega({j_R}+1)](1+\cos\omega)}{\sin\omega(1-\cos\omega)} \\ - 2 ({j_R}+1)\frac{\cos[\omega({j_R}+ 1)]}{(1-\cos\omega)}\Bigr)^2\,.
\end{multline}

In terms of this parametrization, the SU(2) invariant measure is
\begin{equation}
  \mathrm{d}\Omega = \frac{1}{2\pi^2}\sin^2\frac{\omega}{2} \, \sin\theta \, \mathrm{d}\phi \, \mathrm{d}\theta \, \mathrm{d}\omega ,
\end{equation}
where $0\leq \phi < 2\pi$, $0\leq \theta\leq \pi$, and $0\leq\omega\leq\pi$.  For the rotation $U_S(\omega,\theta,\phi)$ on the qubit system in this parametrization, we have
\begin{equation}
  U_S(\omega,\theta,\phi) = \begin{pmatrix}
  \cos\frac{\omega}{2} + i \sin\frac{\omega}{2} \cos\theta & ie^{-i \phi} \sin\frac{\omega}{2} \sin\theta \\
  ie^{i \phi} \sin\frac{\omega}{2} \sin\theta & \cos\frac{\omega}{2} - i \sin\frac{\omega}{2} \cos\theta
  \end{pmatrix}\,.
\end{equation}
It follows that the composition of encoding and decoding maps, given by Eq.~(\ref{eq:Method2}), is
\begin{align} \label{eq:SU2effectivedecoherence}
  \mathcal{R} \circ \mathcal{E}(\rho) &={\textstyle \binom{2{j_R}+3}{3}^{-1}} \int \mathrm{d}\Omega \, |\langle e_{j_R}|(\omega,\theta,\phi)_{j_R}\rangle |^2 \nonumber \\
  &\qquad \qquad \times U_S(\omega,\theta,\phi)\rho U_S(\omega,\theta,\phi)^\dag \nonumber \\
  &=\tfrac{j_R}{{j_R}+1}\rho + \tfrac{1}{{j_R}+1}I/2  \\
  &=\Bigl( \tfrac{j_R}{{j_R}+1}\mathcal{I} + \tfrac{1}{{j_R}+1}\mathcal{G} \Bigr) [\rho ]
  \label{eq:SU(2)Method1}
\end{align}
where $\mathcal{G}$ is the SU(2)-twirling operation (which is completely decohering for a single qubit).

\subsection{Relational subsystems}

Next, we determine the precise nature of the relational subsystems where the quantum information is encoded. In
this example, the multiplicity spaces play a key role. We begin by describing the group-induced structure of the
Hilbert spaces, both for the RF and the total system. Under the action of the representation $U_R$ of SU(2), the
Hilbert space for the reference frame is decomposed as
\begin{equation}
    \label{eq:HdecompSU2RF}
  \mathcal{H}_R = \bigoplus_{j=0}^{j_R} \mathcal{M}^{(j)}_R \otimes \bar{\mathcal{N}}^{(j)}_R \,.
\end{equation}
The joint system $RS$, consisting of the RF plus a spin-1/2 qubit,
carries a collective representation $U_{RS} = U_R \otimes U_S$ of
SU(2) which can easily be determined using standard angular momentum
coupling.  For coupling a spin-$j$ irrep to a spin-1/2 irrep, we
have $\mathcal{M}^{(j)}_R \otimes \mathcal{M}^{(\frac{1}{2})}_S =
\mathcal{M}^{(j+\frac{1}{2})}_{RS} \oplus
\mathcal{M}^{(j-\frac{1}{2})}_{RS}$.  Thus, the Hilbert space of the
joint system $RS$ has a similar decomposition under the action of
$U_{RS}$, given by
\begin{equation}
    \label{eq:HdecompSU2Joint}
  \mathcal{H}_{RS} = \bigoplus_{J=\frac{1}{2}}^{j_R+\frac{1}{2}} \mathcal{M}^{(J)}_{RS} \otimes
  \bar{\mathcal{N}}^{(J)}_{RS} \,.
\end{equation}
The multiplicity spaces for the joint system $RS$ are related to those of the RF as
\begin{equation}
  \label{eq:SU2multspacedirectsum}
  \bar{\mathcal{N}}^{(J)}_{RS} = \begin{cases} \bar{\mathcal{N}}^{(J+\frac{1}{2})}_R \oplus
  \bar{\mathcal{N}}^{(J-\frac{1}{2})}_R \,, & J<j_R+\frac{1}{2} \\
  \bar{\mathcal{N}}^{(j_R)}_{R} \,, & J=j_R+\frac{1}{2}\,.\end{cases}
\end{equation}

For simplicity, we consider the qubit state to be pure, expressed in the standard angular momentum basis as
\begin{equation}
  \label{eq:SpinQubitPure}
  |\psi\rangle = \sum_{s=\pm \frac{1}{2}} \alpha_s |\tfrac{1}{2},s\rangle\,.
\end{equation}
The encoded state within the $J$th irrep is
\begin{equation}
  \Pi^{(J)} (|e_{j_R} \rangle \langle e_{j_R}| \otimes |\psi\rangle \langle \psi|)\Pi^{(J)}\,.
\end{equation}
To evaluate this expression, we first evaluate
\begin{equation}
  \label{eq:JMpurestate}
  \langle J,M| (|e_{j_R}\rangle |\psi\rangle )\,,
\end{equation}
where we recall that $|J,M\rangle$ is defined on the subsystem $\mathcal{M}^{(J)}_{RS}$.  Therefore, the
state~(\ref{eq:JMpurestate}) is an unnormalized vector on $\bar{\mathcal{N}}^{(J)}_{RS}$.  We transform
$|e_{j_R}\rangle |\psi\rangle$ to a coupled basis using Clebsch-Gordan coefficients $(j_1,m_1;j_2,m_2|j,m)$. In
terms of the bases used in (\ref{eq:SpinChiribella}) and (\ref{eq:SpinQubitPure}), we have
\begin{multline}
  |j,m\rangle|\tfrac{1}{2},s\rangle|\phi_{j,m}\rangle \\
  = \sum_{b=\pm \frac{1}{2}} |J{=}j{+}b,M{=}m{+}s\rangle|\phi_{j,m}\rangle (j,m;\tfrac{1}{2},s|j+b,m+s) \,.
\end{multline}
We note that the states $\{ |\phi_{j,m}\rangle\,, m=-j,\dots,j\}$
for $j=J+\tfrac{1}{2}$ ($J-\tfrac{1}{2}$) form a basis of
$\bar{\mathcal{N}}^{(J-\frac{1}{2})}_{R}$
($\bar{\mathcal{N}}^{(J+\frac{1}{2})}_{R}$).  It follows that the
full set of states $\{ |\phi_{j,m}\rangle\,, j=J\pm \tfrac{1}{2},
m=-j,\dots,j\}$ form a basis of $\bar{\mathcal{N}}^{(J)}_{RS}$.  We
have
\begin{align}
  \langle J,M|& (|e_{j_R}\rangle |\psi\rangle ) \nonumber  \\
  &= \sum_{j=0}^{j_R} \sum_{m=-j}^j \sum_{s,b=\pm \frac{1}{2}} \sqrt{\frac{2j+1}{D_R}} \alpha_s |\phi_{j,m}\rangle \nonumber \\
  &\qquad \times (j,m;\tfrac{1}{2},s|j+b,m+s) \delta_{J,j+b}\delta_{M,m+s} \nonumber \\
  &= \sum_{s,b=\pm \frac{1}{2}} \sqrt{\frac{2(J-b)+1}{D_R}} \alpha_s |\phi_{J-b,M-s}\rangle \nonumber \\
  &\qquad \times (J-b,M-s;\tfrac{1}{2},s|J,M) \,.
\end{align}
We use the following Clebsch-Gordan identity,
\begin{multline}
  (j_1,m_1;j_2,m_2|j,m) \\
  = (-1)^{j_2+m_2} \sqrt{\tfrac{2j+1}{2j_1+1}} (j,-m;j_2,m_2|j_1,-m_1) \,,
\end{multline}
to obtain
\begin{align}
  \langle J,M|&(|e_{j_R}\rangle |\psi\rangle ) \nonumber \\
   &= \sqrt{\frac{2J+1}{D_R}} \sum_{s,b=\pm 1/2} \alpha_s (-1)^{s-b+1}  |\phi_{J-b,M-s}\rangle \nonumber \\
   &\qquad \times (J,M;\tfrac{1}{2},-s|J-b,M-s) \,. \label{eq:IntermediateStep}
\end{align}
We now consider two cases for $J$ separately.

For $J<j_R+1/2$, we note that the multiplicity space
$\bar{\mathcal{N}}^{(J)}_{RS}$ is unitarily equivalent to the tensor
product of a spin-$J$ and a spin-1/2 system coupled to total angular
momentum $J\pm 1/2$.  That is,
\begin{equation}
  \label{eq:MagicDecomp}
  \bar{\mathcal{N}}^{(J)}_{RS} = \bar{\mathcal{N}}^{(J+\frac{1}{2})}_R \oplus \bar{\mathcal{N}}^{(J-\frac{1}{2})}_R
  \simeq \mathcal{K}^{(J)} \otimes \mathcal{H}_{S}\,,
\end{equation}
where $\mathcal{K}^{(J)}$ carries an irrep $J$ of SU(2) and $\simeq$ denotes unitary equivalence.  We explicitly
define the bijective isometry $A^{(J)}: \bar{\mathcal{N}}^{(J)}_{RS} \rightarrow \mathcal{K}^{(J)} \otimes
\mathcal{H}_{S}$ via its adjoint action on a basis for $\mathcal{K}^{(J)} \otimes \mathcal{H}_{S}$ as
\begin{align}
\label{eq:intertwineraction}
&A^{(J)\dag}|J,M\rangle_{\mathcal{K}^{(J)}} |\tfrac{1}{2},s\rangle_{\mathcal{H}_S} \nonumber \\
 &= (-1)^{s+\frac{1}{2}} \sum_{b=\pm 1/2} (-1)^{b-\frac{1}{2}}(J,M;\tfrac{1}{2},s|J+b,M+s) \nonumber \\
 &\qquad \qquad\times |\phi_{J+b,M+s}\rangle\,.
\end{align}
In terms of this new subsystem structure for the multiplicity
spaces, we can express (\ref{eq:IntermediateStep}) as
\begin{align}
  \langle J,M| &(|e_{j_R}\rangle |\psi\rangle ) \nonumber \\
  &= \sqrt{\frac{2J+1}{D_R}} \sum_{s=\pm 1/2} \alpha_s A^{(J)\dag}|J,M\rangle_{\mathcal{K}^{(J)}} |\tfrac{1}{2},s\rangle_{\mathcal{H}_S} \nonumber \\
  &= \sqrt{\frac{2J+1}{D_R}} A^{(J)\dag}|J,M\rangle_{\mathcal{K}^{(J)}} |\psi\rangle_{\mathcal{H}_S} \,,
\end{align}
where $|\psi\rangle_{\mathcal{H}_S}$ is defined by
Eq.~(\ref{eq:SpinQubitPure}).
It follows that the
encoded state for an irrep $J$ where $J<j_R+\frac{1}{2}$ is
\begin{align}
  \mathcal{E}^{(J)}(\rho)
  &=\text{Tr}_{\mathcal{M}^{(J)}_{RS}}\bigl[\Pi^{(J)} (|e_{j_R} \rangle \langle e_{j_R}| \otimes \rho )\Pi^{(J)} \bigr] \nonumber \\
  &= \sum_{M=-J}^J \langle J,M| (|e_{j_R} \rangle \langle e_{j_R}| \otimes \rho)|J,M\rangle \nonumber \\
  &= \frac{2J+1}{D_R} A^{(J)\dag}\bigl( I_{\mathcal{K}^{(J)}} \otimes \rho \bigr) A^{(J)} \,, \quad J<j_R+\tfrac{1}{2}\,.
  \label{eq:UsualJqubit}
\end{align}
Because $A^{(J)}$ is bijective, (specifically, because the set of
states $\{A^{(J)\dag}|J,M\rangle_{\mathcal{K}^{(J)}}
|\tfrac{1}{2},s\rangle_{\mathcal{H}_S}\,; M=-J,\ldots,J \,, \
s=\pm\tfrac{1}{2}\}$ are orthogonal) we find the qubit faithfully
encoded into the relational subsystem whenever $J < j_R +
\frac{1}{2}$.

However, for the result $J = j_R + \frac{1}{2}$, the multiplicity
space $\bar{\mathcal{N}}^{(J)}_{RS}$ is exceptional;
see~(\ref{eq:SU2multspacedirectsum}).   We cannot factorize
$\bar{\mathcal{N}}^{(J)}_{RS}$ as in Eq.~(\ref{eq:MagicDecomp}).
Nonetheless, we can still introduce a Hilbert space
$\mathcal{K}^{(j_R+\frac{1}{2})}$ which carries an irrep
$j_R+\tfrac{1}{2}$ of SU(2) and in terms of it we can define an
isometry $A^{(j_R+\frac{1}{2})}:
\bar{\mathcal{N}}^{(j_R+\frac{1}{2})}_{RS} \rightarrow
\mathcal{K}^{(j_R+\frac{1}{2})} \otimes \mathcal{H}_{S}$, by
modifying Eq.~(\ref{eq:intertwineraction}) to include only the
$b=-\tfrac{1}{2}$ term in the sum.  This isometry is simply not
surjective.  It follows that the set of states
$\{A^{(j_R+\frac{1}{2})\dag}|j_R+\tfrac{1}{2},M\rangle_{\mathcal{K}^{(j_R+\frac{1}{2})}}
|\tfrac{1}{2},s\rangle_{\mathcal{H}_S}\,;
M=-j_R-\tfrac{1}{2},\ldots,j_R+\tfrac{1}{2} \,, \
s=\pm\tfrac{1}{2}\}$ are no longer orthogonal. Therefore, the map
\begin{equation}
  \label{eq:ExceptionalJqubit}
  \mathcal{E}^{(j_R+\frac{1}{2})}(\rho)
  = \frac{2j_R+2}{D_R} A^{(j_R+\frac{1}{2})\dag}\bigl( I_{\mathcal{K}^{(j_R+\frac{1}{2})}}
  \otimes \rho \bigr) A^{(j_R+\frac{1}{2})}\,,
\end{equation}
is no longer invertible.  The action of $A^{(j_R+\frac{1}{2})\dag}$
can be viewed as a projection of uncoupled states on
$\mathcal{K}^{(j_R+\frac{1}{2})} \otimes \mathcal{H}_{S}$ onto the
subspace of states which couple to total angular momentum $j_R$.

The probability assigned to each irrep $J$ is
\begin{align} \label{eq:probsCartesian}
  p_J &= \text{Tr}\bigl[\Pi^{(J)} (|e_{j_R} \rangle \langle e_{j_R}| \otimes |\psi\rangle \langle \psi|)\Pi^{(J)} \bigr] \nonumber \\
  &= \begin{cases} \frac{(2J+1)^2}{D_R} \,, & J<j_R+\frac{1}{2}\,, \\
  \frac{(2j_R+1)(j_R+1)}{D_R} \,, & J=j_R+\frac{1}{2}\,. \end{cases}
\end{align}
We note that these probabilities satisfy $\sum_{J=\frac{1}{2}}^{j_R+\frac{1}{2}} p_J = 1$.

The decoding map within each irrep $J$ takes the form
\begin{align}
  \mathcal{R}^{(J)}(\cdot) &= \text{Tr}_{\mathcal{K}^{(J)}} [A^{(J)}
  (\cdot) A^{(J)\dag} ] \nonumber \\
  &=(2J+1) \langle e_{j_R}| I_{\mathcal{M}^{(J)}_{RS}} \otimes (\cdot) |e_{j_R}\rangle \,.
\end{align}
For $J<j_R+\frac{1}{2}$, Eq.~(\ref{eq:UsualJqubit}) gives
\begin{equation}
  (\mathcal{R}^{(J)}\circ \mathcal{E}^{(J)})(\rho) = \rho \,.
\end{equation}
For $J=j_R+\frac{1}{2}$, we make use of Eq.~(\ref{eq:ExceptionalJqubit}) and find
\begin{multline}
  (\mathcal{R}^{(j_R+\frac{1}{2})} \circ \mathcal{E}^{(j_R+\frac{1}{2})})(\rho) \\
  = \tfrac{(2j_R+1)(j_R+1)}{D_R} \Bigl( \tfrac{2j_R}{6({j_R}+1)}\mathcal{I} + \tfrac{(2j_R+3)}{3({j_R}+1)}\mathcal{G} \Bigr) \bigl[\rho \bigr]\,,
\end{multline}
where $\mathcal{G}$ is the SU(2)-twirling operation (complete decoherence).  Averaging over the irreps with the
weights given in Eq.~(\ref{eq:probsCartesian}), we find the decoded state to be
\begin{align}
  (\mathcal{R}\circ \mathcal{E})(\rho) &= \sum_{J=\frac{1}{2}}^{j_R-\frac{1}{2}} \tfrac{(2J+1)^2}{D_R} \mathcal{I}[\rho]\nonumber \\
  &\qquad + \tfrac{(2j_R+1)(j_R+1)}{D_R} \Bigl( \tfrac{2j_R}{6({j_R}+1)}\mathcal{I} + \tfrac{(2j_R+3)}{3({j_R}+1)}\mathcal{G} \Bigr) [\rho] \nonumber \\
  &=\Bigl( \tfrac{j_R}{{j_R}+1}\mathcal{I} + \tfrac{1}{{j_R}+1}\mathcal{G} \Bigr) [\rho ]\,,
\end{align}
thereby verifying that Eq.~(\ref{eq:SU2effectivedecoherence}) is recovered.

We note that our recovery operation is perfect in the
non-exceptional irreps and therefore optimal there. In the
exceptional irrep, the error incurred in recovery is at most twice
that of the optimal recovery, as described in
Sec.~\ref{sec:decoding}, by virtue of being of the form of Barnum
and Knill's approximate reversal.  We leave it as an open problem to
to prove whether this recovery is in fact optimal, or if not, to
identify what form an optimal recovery map would take.

Post-selectively perfect decoding is achieved if Bob implements a projective measurement that
distinguishes the irreps and obtains $J \ne j_R+\frac{1}{2}$.  By Eq.~(\ref{eq:probsCartesian}) and
Eq.~(\ref{eq:DRCartesian}) (or by Eq.~(\ref{eq:pperfect}) directly), we see that the probability for this to
occur is
\begin{align}
 p_{\textrm{perfect}} &= \frac{1}{D_R} \sum_{J=\frac{1}{2}}^{j_R-\frac{1}{2}} (2J+1)^2
 = \frac{\sum_{J=\frac{1}{2}}^{j_R-\frac{1}{2}} (2J+1)^2}{\sum_{j=0}^{j_R} (2j+1)^2} \nonumber \\
 &= \frac{2j_R}{2j_R + 3} \,.
\end{align}

\section{Example: Direction indicator}
\label{sec:Direction}

A directional RF identifies only a single direction in space, as
opposed to a full set of axes. Such an RF is not associated directly
with a Lie group, but rather with a coset space.  Specifically,
although SU(2) may provide a group of transformations between all
possible directional RFs, any one directional RF is invariant under
U(1) rotations about its axis of symmetry; the relevant coset is
then SU(2)/U(1).

Because of this distinction, we expect this example to proceed differently from the other two. The distinction
is immediately apparent because there is no obvious candidate for a fiducial reference state as in
Eq.~(\ref{eq:Chiribella}).  Instead, we take the directional RF to be in an SU(2)-coherent state of size $j_R$,
so that the fiducial state is
\begin{equation} \label{eq:fiducialdirection}
  |e_{j_R}\rangle = |{j_R},m={j_R}\rangle \,.
\end{equation}
We consider a qubit system which is described relative to Alice's local Cartesian frame by the state
\begin{equation}
  \rho =\left\vert \psi \right\rangle \left\langle \psi \right\vert \text{
  where }\left\vert \psi \right\rangle ={\alpha \left\vert 0\right\rangle
  +\beta \left\vert 1\right\rangle },
\end{equation}
where $\left\vert 0(1)\right\rangle =\left\vert {j_R}=1/2,m=\pm 1/2\right\rangle$. Note first of all that even
if Bob shared a classical reference direction with Alice, her $\hat{z}$ axis for instance, his description of
the system is still related to Alice's by a dephasing operation. The reason is that he \emph{only} shares
Alice's $\hat{z}$-axis, and so the rotation about $\hat{z}$ that relates his local $\hat{x}$-axis to hers is
completely unknown.  Averaging over rotations $R_z(\theta) = \exp(-i\theta J_z)$ yields the dephasing operation
\begin{equation}
  \label{eq:Dephasing}
  \mathcal{U}[\rho] = \int_0^{2\pi} \frac{d\theta}{2\pi} R_z(\theta) \rho R^\dag_z(\theta) \\
= |\alpha|^2 |0\rangle \langle 0| + |\beta|^2 |1\rangle \langle 1|\,.
\end{equation}
Consequently, if Bob has a bounded-size token of Alice's $\hat{z}$-axis, his decoding will yield a state that
approaches $\mathcal{U}(\rho)$ rather than $\rho$ as one increases the size of the token.

\subsection{Effective decoherence}

Define $|\Omega_{j_R}\rangle = U_R(\Omega)|e_{j_R}\rangle$ where $\Omega \in \textrm{SU(2)}$.  The encoding of
$\rho$ relative to Bob's local Cartesian frame is the following state of the composite of RF token and system
\begin{equation} \label{eq:encodingdirection}
  \mathcal{E}(\rho)=\int d\Omega |\Omega_{j_R} \rangle\langle \Omega_{j_R}| \otimes U_S(\Omega)\rho U_S^\dag(\Omega) \,.
\end{equation}
To decode, Bob measures the covariant POVM $\{D_{R} |\Omega'_{j_R}\rangle \langle \Omega'_{j_R}| d\Omega' \}$ on
the RF token and reorients the system accordingly.  The net result is
\begin{equation} \label{eq:decodingdirection}
  \mathcal{R}\circ \mathcal{E}(\rho) = D_{R}\int d\Omega |\langle e_{j_R}|\Omega_{j_R}\rangle|^{2} U_S(\Omega)\rho U_S^{\dag }(\Omega)\,.
\end{equation}
The effect of this map will be particularly simple given that $|e_{j_R}\rangle$ is invariant under U(1) rotations about the $z$-axis.

For this calculation, it will be easiest to use Euler angles to parametrize SU(2):
\begin{equation}
  U(\Omega) = e^{-ia J_z} e^{-ib J_y} e^{-ic J_z} \,,
\end{equation}
where $a,b,c \in [0,2\pi]$.  (We note this parametrization is
different from that used in Sec.~\ref{sec:Cartesian}.) With this
parametrization,
\begin{align}
  \langle e_{j_R}|\Omega_{j_R} \rangle  &= \langle {j_R},{j_R}| U_R^{(j_R)}(\Omega)|{j_R},{j_R}\rangle \nonumber \\
  &=e^{-i(a+c) {j_R}} [\cos(b/2)]^{2{j_R}}\,,
\end{align}
and thus $|\langle e_{j_R}|\Omega_{j_R} \rangle |^{2} =
[\cos(b/2)]^{4{j_R}}$.  We can express $U_S(\Omega)$ as a $2\times
2$ matrix in the $z$-basis as
\begin{equation}
  U_S(\Omega) = \begin{pmatrix} e^{-i\frac{a}{2}} & 0 \\ 0 & e^{i\frac{a}{2}}
  \end{pmatrix} \begin{pmatrix} \cos b/2 & -\sin b/2  \\ \sin b/2
  & \cos b/2 \end{pmatrix} \begin{pmatrix} e^{-i\frac{c}{2}} & 0 \\ 0 & e^{i\frac{c}{2}}
  \end{pmatrix} \,,
\end{equation}
and our qubit system in Bloch vector notation as
\begin{equation}
  \rho = \frac{1}{2}\begin{pmatrix} 1+ z & x-iy \\ x+iy & 1-z
  \end{pmatrix} \,.
\end{equation}
Using the identities
\begin{align}
  \int_0^\pi \mathrm{d}b\,\sin b\,\cos^{4{j_R}}(b/2) &= \frac{2}{2{j_R}+1}\,,\\
  \int_0^\pi \mathrm{d}b\,\sin b\,\cos^{4{j_R}}(b/2)\,\cos^2(b/2)
  &= \frac{1}{{j_R}+1}\,,
\end{align}
we find that
\begin{align} \label{eq:effectivedecoherencedirection}
  \mathcal{R} \circ \mathcal{E}(\rho)
  &=(2{j_R}+1)\int\mathrm{d}\Omega \, \cos^{4{j_R}}(b/2)  \, U_S(\Omega) \rho U_S(\Omega)^{-1} \nonumber \\
  &= \frac{1}{2}\begin{pmatrix} 1+\frac{j_R}{j_R+1}z & 0 \\ 0 &
  1-\frac{j_R}{j_R+1}z \end{pmatrix} \nonumber \\
  &= \Bigl( \bigl( \tfrac{{j_R}}{{j_R}+1}\mathcal{I} + \tfrac{1}{{j_R}+1}\mathcal{G} \bigr)\circ \mathcal{U} \Bigr) [\rho ] \,,
\end{align}
where $\mathcal{G}$ is the SU(2)-twirling operation, and $\mathcal{U}$ is the dephasing operator defined in
Eq.~(\ref{eq:Dephasing}).

\subsection{Relational subsystems}

First, we note that the RF token, consisting of only a spin-$j_R$
system, does not possess a multiplicity space. When coupling this
spin-$j_R$ system to the spin-1/2 qubit, the resulting collective
system is described by
\begin{equation} \label{eq:Hdecomp_direction}
  \mathcal{H}_{RS} = \mathcal{M}^{(j_R)}_R \otimes \mathcal{M}^{(\frac{1}{2})}_S = \mathcal{M}^{(j_R+\frac{1}{2})}_{RS} \oplus \mathcal{M}^{(j_R-\frac{1}{2})}_{RS}\,,
\end{equation}
and does not possess any multiplicity spaces either.

Given that the fiducial state, Eq.~(\ref{eq:fiducialdirection}), is not of the form of
Eq.~(\ref{eq:Chiribella}), the derivation of Eq.~(\ref{eq:Edecomp1}) is no longer valid.  Nonetheless, the
encoding map defined by Eq.~(\ref{eq:encodingdirection}) may still be written in the form of
Eq.~(\ref{eq:Edecomp1}), namely,
\begin{equation} \label{eq:qqq}
  \mathcal{E}(\rho) = \sum_{J=j_R-\frac{1}{2}}^{j_R+\frac{1}{2}} d_J^{-1} I_{\mathcal{M}_{RS}^{(J)}} \otimes
  \mathcal{E}^{(J)}(\rho)\,,
\end{equation}
where
\begin{equation} \label{eq:qqq2}
  \mathcal{E}^{(J)}(\rho) =\text{Tr}_{\mathcal{M}^{(J)}_{RS}}\bigl[\Pi^{(J)} (|e_{j_R} \rangle \langle e_{j_R}|
  \otimes \rho )\Pi^{(J)} \bigr]\,,
\end{equation}
which is of the same form as Eq.~(\ref{eq:Equsefulform}).  To see that this decomposition exists, simply express
Eq.~(\ref{eq:encodingdirection}) as $\mathcal{E}(\rho)=\mathcal{G}(|e_{j_R}\rangle \langle e_{j_R}|\otimes
\rho)$ where $\mathcal{G}$ is the SU(2)-twirling operation.  Then, using
Eq.~(\ref{eq:GeneralDecoherenceDfullDfree}) and Eq.~(\ref{eq:Hdecomp_direction}), we have
\begin{align}
  \mathcal{E}(\rho) = \sum_{J=j_R-\frac{1}{2}}^{j_R+\frac{1}{2}} \mathcal{D}_{\mathcal{M}_{RS}^{(J)}} \bigl[\Pi^{(J)} (|e_{j_R} \rangle \langle e_{j_R}|
  \otimes \rho )\Pi^{(J)} \bigr] \,,
\end{align}
which is equivalent to Eqs.~(\ref{eq:qqq}) and~(\ref{eq:qqq2}). Note that $\mathcal{E}^{(J)}$ is still a map
from $\mathcal{B}(\mathcal{H}_S)$ to $\mathcal{N}_{RS}^{(J)}$, but in this case
$\mathcal{N}_{RS}^{(J)}=\mathbb{C}$, so that it maps density operators to scalars.  Specifically,
\begin{align} \label{eq:EJexplicit}
  \mathcal{E}^{(J)}(\rho) &=
  \begin{cases}
   \frac{2j_R}{2j_R+1} |\alpha |^{2} + \frac{1}{2j_R+1} \,, & J=j_R + \frac{1}{2}\,, \\
  \frac{2{j_R}}{2{j_R}+1} |\beta |^{2} \,, & J=j_R - \frac{1}{2}\,.
  \end{cases}
\end{align}
We see that the encoding operation in this case can be described as
follows: after adjoining the RF token to the system, destructively
measure the total angular momentum $\bold{J}^2$ of the composite and
upon obtaining quantum number $J$, reprepare the system in the
associated SU(2)-invariant state $d_J^{-1}
I_{\mathcal{M}_{RS}^{(J)}}$.


The decoding operation defined in Eq.~(\ref{eq:decodingdirection}) is clearly proportional to the
Hilbert-Schmidt adjoint of the encoding operation.  Consequently, it admits a decomposition into irreps via the
adjoints of Eq.~(\ref{eq:qqq}), namely,
\begin{equation}
  \mathcal{R}(\rho_{RS})= \sum_{J=j_R-\frac{1}{2}}^{j_R+\frac{1}{2}} \mathcal{R}^{(J)}\bigl[
  \textrm{Tr}_{\mathcal{M}_{RS}^{(J)}}\bigl( \Pi^{(J)} \rho_{RS} \Pi^{(J)} \bigr) \bigr]
  \,,
\end{equation}
where $\mathcal{R}^{(J)} \propto \mathcal{E}^{(J)\dag}$ is a map from $\mathbb{C}$ to
$\mathcal{B}(\mathcal{H}_S)$. This is of the same form as Eq.~(\ref{eq:decodingdecomposition}).  To determine
$\mathcal{R}^{(J)}$, we calculate the adjoint of Eq.~(\ref{eq:qqq2}) and determine the normalization by
requiring that $\text{Tr}[\mathcal{R}^{(J)}(1)]=1$ for all $J$.
We obtain
\begin{equation}
  \mathcal{R}^{(J)}(p)=\frac{2j_R+1}{2J+1} \langle e_{j_R}| I_{\mathcal{M}_{RS}^{(J)}}|e_{j_R}\rangle \times p \,,
\end{equation}
where $p\in \mathbb{C}$. Except for the constant of proportionality, this has the form of
Eq.~(\ref{eq:Rqusefulform}).  It evaluates to
\begin{align} \label{eq:RJp}
  \mathcal{R}^{(J)}(p) &= p\times
  \begin{cases}
 \frac{2j_R+1}{2j_R+2} |0\rangle\langle 0| + \frac{1}{2j_R+2}|1\rangle\langle 1|  \,, & J=j_R + \frac{1}{2}\,, \\
 |1\rangle\langle 1| \,, & J=j_R - \frac{1}{2}
  \end{cases}
\end{align}
[which one could also infer by taking the adjoint of Eq.~(\ref{eq:EJexplicit})]. Consequently, the decoding
operation may be described as follows: destructively measure the total angular momentum (squared), $\bold{J}^2$,
on the composite of RF token and system and upon obtaining quantum number $j_R\pm\frac{1}{2}$, reprepare the
system in one of the two states in Eq.~(\ref{eq:RJp}).

The composition of encoding and decoding yields
\begin{align}
  \mathcal{R} \circ \mathcal{E}(\rho)
&= \sum_{J=j_R-\frac{1}{2}}^{j_R+\frac{1}{2}} \mathcal{R}^{(J)} \circ \mathcal{E}^{(J)}(\rho) \nonumber \\
  &=\frac{1}{{j_R}+1}\bigl( \frac{1}{2}|0\rangle \langle 0| + \frac{1}{2}|1\rangle\langle 1| \bigr) \nonumber \\
  &\qquad + \frac{{j_R}}{{j_R}+1}\bigl( |\alpha |^{2}|0\rangle \langle 0| +|\beta |^{2}|1\rangle \langle 1| \bigr) \nonumber \\
  &= \Bigl( \bigl( \tfrac{{j_R}}{{j_R}+1}\mathcal{I} + \tfrac{1}{{j_R}+1}\mathcal{G} \bigr)\circ \mathcal{U} \Bigr) [\rho ] \,,
\end{align}
in agreement with Eq.~(\ref{eq:effectivedecoherencedirection}). We note that this coincides with the result
obtained by Poulin~\cite{Pou06}.

This recovery operation is of the form of the approximate reversal
operation described by Barnum and Knill~\cite{Bar02} and is
therefore near-optimal in the sense described in
Sec.~\ref{sec:decoding}. Although it is not itself the optimal
recovery map, the latter is easy to find in this example and we do
so presently.

Given that the only pure states of the system that one can hope to reconstruct in the limit of an unbounded RF
token are the $J_z$ eigenstates, denoted here by $|0\rangle \langle 0|$ and $|1\rangle \langle 1|$, it is
natural to take as our figure of merit the average input-output fidelity equally weighted over these two input
states (because then one can achieve fidelity 1 in the limit of an unbounded RF token). The only information
about the state of the system that is encoded in $\mathcal{E}(\rho)$ is encoded in the relative weights of its
two SU(2)-invariant terms. Consequently, the optimal decoding operation must consist of a determination of $J$
followed by a repreparation of the system state.

We make use of previous work on the optimal estimation of the relative angle between a spin-1/2 system and a
spin-$j_R$ reference frame \cite{BRS04b}. These results show that a measurement of $\bold{J}^2$ on the composite
is in fact optimal for estimating whether the system state was $|0\rangle \langle 0|$ or $|1\rangle \langle 1|$
given a uniform prior over the two. It is also shown there that the posterior probabilities one ought to assign
to these two states upon obtaining outcomes $j_R + \frac{1}{2}$ and $j_R - \frac{1}{2}$ are
\begin{align}
  p(0|+)&=\frac{2j+1}{2j+2}\,, \quad  &p(1|+) &= \frac{1}{2j+2} \,, \nonumber \\
  p(0|-)&=0\,, &p(1|-)&=1 \,.
\end{align}
In order to optimize the fidelity, one must reprepare the state that
is most likely given the outcome, so that one should reprepare
$|0\rangle \langle 0|$ given the `$+$' outcome and $|1\rangle
\langle 1|$ given the `$-$' outcome. $\mathcal{R}$ falls short of
this optimal recovery because, by Eq.~(\ref{eq:RJp}), it does not
reprepare $|0\rangle \langle 0|$ given the `$+$' outcome; instead,
it prepares a mixed state reflecting Bob's knowledge of the state
given the measurement outcome.

Finally, we note that there is no possibility for post-selectively perfect recovery of $\mathcal{U}(\rho)$ from
$\mathcal{E}(\rho)$. Both irreps, $j_R \pm 1/2$, encode the state of the system imperfectly.

\begin{acknowledgments}
We thank Robin Blume-Kohout, J.-C. Boileau, Matthias Christandl and Renato Renner for helpful discussions. SDB
acknowledges the support of the Australian Research Council. RWS acknowledges support from the Royal Society.
PST acknowledges the support of the AIF, iCORE, MITACS, and JSPS.
\end{acknowledgments}

\end{document}